\documentclass[draft]{agujournal2019}
\usepackage{url} 
\usepackage{soul}
\usepackage{amsmath}
\usepackage{amssymb}
\usepackage{color}
\linenumbers
\draftfalse

\newcommand{\cor}[1]{{#1}}

\newcommand{\la}{\left<}
\newcommand{\ra}{\right>}

\newcommand{\bnabla}{\boldsymbol{\nabla}}

\def\lb0{\ell_{*, \text{vg}}}
\def\Db0{{D_{*,\text{vg}}}}

\journalname{AGU Advances}

\nolinenumbers

\begin{document}

%
%


\title{A quantitative scaling theory for meridional heat transport in planetary atmospheres and oceans}

%
%




\authors{Basile Gallet\affil{1}, and Raffaele Ferrari\affil{2}}

\affiliation{1}{Universit\'e Paris-Saclay, CNRS, CEA, Service de Physique de l'Etat Condens\'e, 91191 Gif-sur-Yvette, France.}
\affiliation{2}{Massachusetts Institute of Technology, USA}





\correspondingauthor{Basile Gallet}{basile.gallet@cea.fr}




\begin{keypoints}
\item A recent scaling theory for the turbulent heat transport by vortices is extended to include the impact of planetary curvature, in the framework of the two-layer quasigeostrophic beta-plane model.
\item The theory leads to a quantitative parameterization providing the meridional temperature profile in terms of the externally imposed heat flux.
\item The theory provides a quantitative prediction for the emergent criticality, i.e., the degree of instability in a canonical model of planetary atmosphere.
\end{keypoints}

%
%

%
%


\begin{abstract}
The meridional temperature profile of the upper layers of planetary atmospheres is set through a balance between differential radiative heating by a nearby star, or by intrinsic heat fluxes emanating from the deep interior, and the redistribution of that heat across latitudes by turbulent flows. These flows spontaneously arise through baroclinic instability of the meridional temperature gradients maintained by the forcing. When planetary curvature is neglected, this turbulence takes the form of coherent vortices that mix the meridional temperature profiles. However, the curvature of the planet favors the emergence of Rossby waves and zonal jets that restrict the meridional wandering of the fluid columns, thereby reducing the mixing efficiency across latitudes. A similar situation arises in the ocean, where the baroclinic instability of zonal currents leads to enhanced meridional heat transport by a turbulent flow consisting of vortices and zonal jets. A recent scaling theory for the turbulent heat transport by vortices is extended to include the impact of planetary curvature, in the framework of the two-layer quasigeostrophic beta-plane model. This leads to a quantitative parameterization providing the meridional temperature profile in terms of the externally imposed heat flux in an idealized model of planetary atmospheres and oceans. In addition, it provides a quantitative prediction for the emergent criticality, i.e., the degree of instability in a canonical model of planetary atmosphere or ocean.
\end{abstract}

\section*{Plain Language Summary}
The turbulent motion of planetary atmospheres and oceans is greatly impacted by the curvature of the planet, which favors the emergence of Rossby waves and sharp zonal jets. These structures suppress meridional heat transport and impact the resulting thermal structure of the atmosphere or ocean. A quantitative scaling-theory is derived that takes planetary curvature into account when predicting the turbulent heat transport. The theory provides a quantitative parameterization to determine the meridional temperature profile in a canonical model of planetary atmosphere or ocean subject to meridionally dependent heating.


\section{Introduction}

\begin{figure}[t]
\centering
\hspace{-0.5 cm}
\includegraphics[width=10 cm]{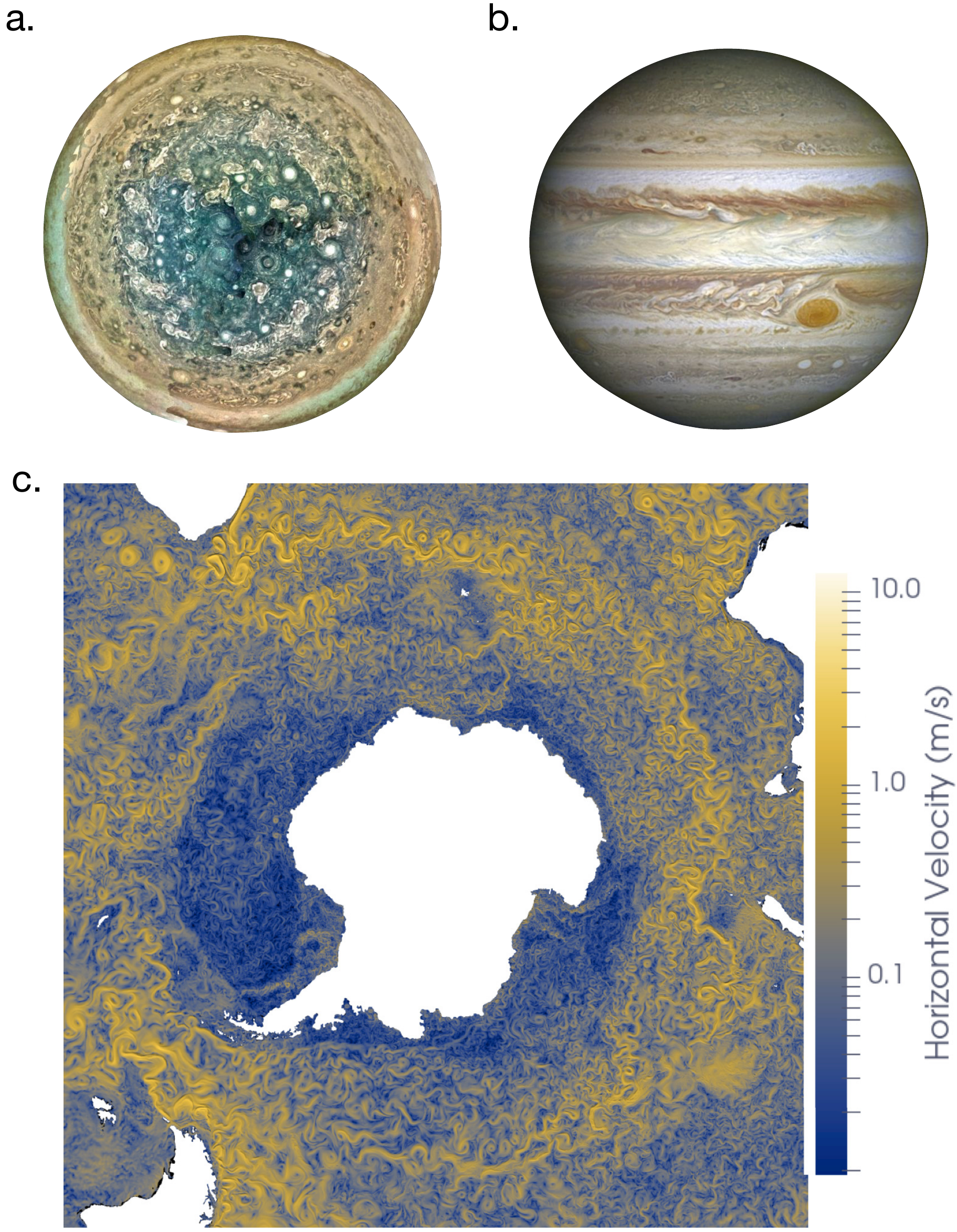}
\caption{\textbf{Baroclinic turbulence in planetary atmospheres and oceans. a:} South pole of Jupiter (composite image from NASA Juno mission). Tubulence takes the form of isolated vortices. {\bf b:} Side view of Jupiter, showing the coherent zonal jets at mid-latitudes (Hubble space telescope).  {\bf c:} \cor{Magnitude of the horizontal velocity 20 meters below the surface from a numerical simulation of the Southern Ocean run at 2.5-km resolution by the ICON-N group~\cite{Korn2017}. The flow field is characterized by turbulent eddies and jets, coherent eddies being more visible in plots of vorticity.} \label{fig:SO_Jupiter}}
\end{figure}

The external fluid layers of planets and their satellites are subject to meridionally dependent heating due to incoming radiation from a distant star, or intrinsic heat fluxes emanating from the planetary interior. On a rocky planet without an atmosphere, such a heat source would induce a strong difference in surface temperature between the equator and the poles. The presence of an atmosphere and/or an ocean strongly mitigates that temperature difference: the meridional temperature gradient induces turbulence in these external fluid layers through a process called baroclinic instability. This baroclinic turbulence, much like a giant stirring spoon or convection in a heated pot of water, greatly enhances heat transport from the equator to the poles, thus reducing the emergent meridional temperature gradient. Predicting the equilibrated meridional temperature profile of these external fluid layers is arguably one of the central questions that a theory of climate should address. As shown in Figure~\ref{fig:SO_Jupiter}, baroclinic turbulence is visible on the surface of Jupiter, where it takes the form of isolated vortices in the polar regions (panel~\ref{fig:SO_Jupiter}a) while inducing coherent zonal jets at mid-latitude (panel~\ref{fig:SO_Jupiter}b). This coexistence of jets and vortices also arises in the Southern Ocean, where baroclinic turbulence results from the instability of the meridional temperature gradient associated with the Antarctic Circumpolar Current: in the near-surface velocity map of panel~\ref{fig:SO_Jupiter}c, ring-shaped structures correspond to isolated vortices, while the zonally elongated features correspond to zonal jets.

The images in Figure~\ref{fig:SO_Jupiter} show that the equilibrated state of baroclinic turbulence consists of a turbulent flow whose energy containing scale -- roughly estimated as the inter-vortex or inter-jet distance -- is small compared to the size of the planet or ocean basin, which is also the extent of the large-scale heating pattern: scale separation spontaneously emerges in this problem, which opens an avenue to describe the turbulent heat transport in terms of a diffusive closure. A diffusive closure relates the turbulent heat flux to the meridional temperature gradient through a diffusivity coefficient that encodes the macroscopic transport induced by the small-scale erratic turbulent motion. The situation is analogous to the textbook examples of molecular diffusion, where a macroscopic downgradient flux of heat or particles is induced by molecular agitation at microscopic scale. In both cases the role of theory is to express the diffusion coefficient in terms of external parameters~\cite{Einstein,Held99}. That scale separation spontaneously arises in a fully turbulent flow is more the exception than the rule, and the reader accustomed to the standard textbook examples of turbulent flows -- Couette flow, Rayleigh-B\'enard convection, etc. -- which all lack scale separation, ought to be skeptical about the use of a diffusive closure. That reader would be right in the sense that the diffusive closure is valid only when applied at scales larger than the energy containing scale, but this is precisely what a parameterization of turbulence in coarse atmosphere and/or ocean models intends to do: the diffusive closure is to be used as a parameterization of turbulent transport in a coarse model that cannot resolve the turbulent eddies and jets that populate baroclinic turbulence. On the other hand, the diffusive approach does not hold at the scale of a single eddy or jet.

\begin{figure}[t]
\includegraphics[width=14 cm]{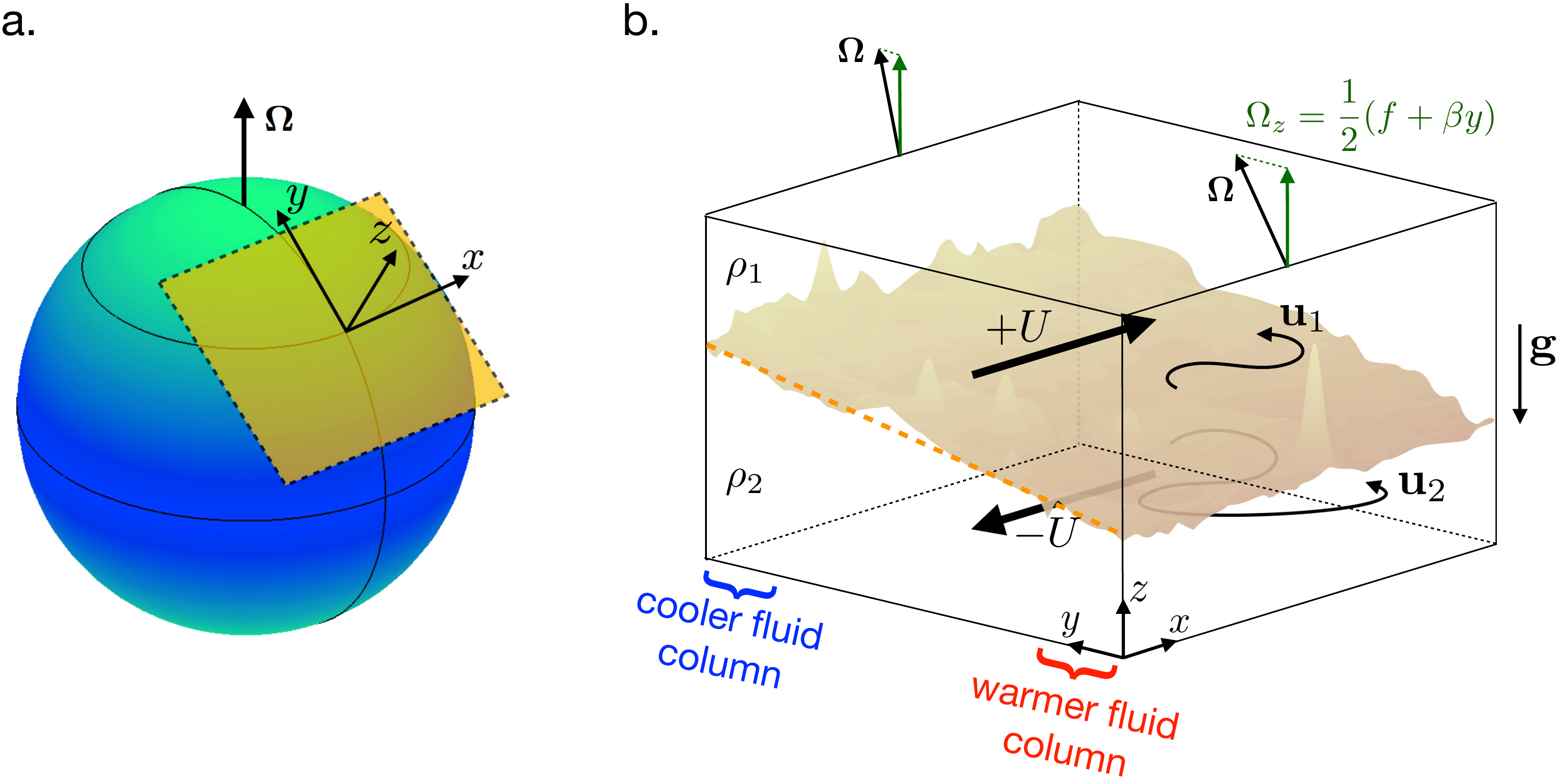} 
\caption{{\bf a:} One can isolate a patch of atmosphere or ocean and study its dynamics in a local Cartesian frame attached to the plane tangent to the sphere at that location. {\bf b:} 3D schematic of the 2LQG system with imposed vertical shear (thick horizontal arrows). The interface is tilted in the $y$ direction as a consequence of thermal wind balance (mean tilt shown with dashed orange line). On vertical average, the fluid is cooler in the Northern region of the domain (i.e., for larger $y$). The Coriolis parameter is equal to twice the projection $\Omega_z$ of the planetary rotation vector $\boldsymbol{\Omega}$ onto the local vertical. It varies linearly with $y$ in the $\beta$-plane approximation, a consequence of the curvature of the planet in this Cartesian-domain approximation. \label{fig:schematic}}
\end{figure}

As explained in \citeA{Pavan} and \citeA{Held99}, one can leverage the spontaneous scale separation to drastically simplify to the study of baroclinic turbulence: instead of running costly spherical models for a whole planet or ocean, one can study patches of atmosphere or ocean of intermediate size -- larger than the inter-jet and inter-vortex distance, but much smaller than the planetary radius -- in isolation, with the goal of determining the local turbulent heat flux, or the local turbulent heat diffusivity, as a function of the local control parameters. The diffusive closure can then be used to represent the heat transport driven by baroclinic turbulence in coarse models of planetary atmospheres and oceans.

A few additional approximations can be made to describe the dynamics of baroclinic turbulence. First, large-scale atmospheric and ocean flows in stratified and rapidly rotating fluids generate structures whose horizontal extent is much larger than their vertical extent, like the vortices and jets in Figure~\ref{fig:SO_Jupiter}. Vertical motions are suppressed, and horizontal motions are set through a near balance between the Coriolis acceleration and the horizontal pressure gradients. Such dynamics are captured by the quasi-geostrophic (QG) equations, an asymptotic expansion of the complete fluid dynamics equations that filters out small and fast scale motions, like gravity waves~\cite{Salmonbook,Vallisbook}. The QG equations will therefore be used in this study. Second, by restricting the analysis to a small patch of atmosphere or ocean, the dynamics can be described in the plane tangent to the sphere at the location of the patch, adopting a local Cartesian frame with $x$ along the zonal direction, $y$ along the meridional one, and $z$ upwards along the local vertical (see panel~\ref{fig:schematic}a). Third, for a patch size much smaller than the planetary radius, the planetary rotation enters the fluid dynamics of  shallow stratified fluids only through its projection onto the local vertical axis $z$: this is the $f$-plane approximation, where $f$ refers to the Coriolis parameter, equal to twice the projection of the rotation vector onto the local vertical. This approximation typically captures the vortical structures of the turbulent flow visible in panel~\ref{fig:SO_Jupiter}a, but it misses the jet-like features visible in panels~\ref{fig:SO_Jupiter}b and \ref{fig:SO_Jupiter}c. A crucial improvement over the $f$-plane model consists in including the local variation of the Coriolis parameter with the North-South coordinate $y$: Taylor expansion in $y$ yields the local Coriolis parameter $f+\beta y$, where $f$ and $\beta$ are constant coefficients. The resulting local framework, referred to as the $\beta$-plane, will be adopted here to study the impact of planetary curvature (through the coefficient $\beta$) within a Cartesian coordinate system.

Focusing on two-dimensional vertically invariant flows on the $\beta$-plane, Rhines first demonstrated that the latitudinal variation in Coriolis parameter has a profound impact on the ensuing turbulence~\cite{Rhines75}. In the absence of planetary curvature ($\beta=0$), the two-dimensional turbulence generated by any random forcing results in the spontaneous emergence of ever growing coherent flow structures through a process referred to as an inverse energy cascade (as opposed to the spontaneous breakup of coherent structures into smaller ones characteristic of three dimensional turbulence in the absence of stratification and rotation). The addition of $\beta$ supports a new class of waves, the Rossby waves. Rossby-wave interactions at large scales channel the energy into increasingly elongated structures in the zonal direction and zonal jets eventually emerge, with reduced meridional velocity fluctuations and thus reduced meridional heat transport. This phenomenology has been proposed to explain both the banded jets observed in gaseous planets like Jupiter~\cite{Ingersoll,Kaspi} and the jet-like flow features observed in the Southern Ocean~\cite{Thompson10}, see Figure~\ref{fig:SO_Jupiter}.

The reduction of meridional transport by $\beta$ can also be understood from the point of view of baroclinic instability. While baroclinic instability on the $f$-plane arises for arbitrarily weak meridional temperature gradient, $\beta$ stabilizes weak meridional temperature gradients in the simplest setting of (undamped) baroclinic turbulence, introduced in section~\ref{sec:Homogeneous}. The consequence is a nonzero threshold value of the temperature gradient below which the fluid remains motionless, and above which baroclinic turbulence arises. This observation led to the `baroclinic adjustment' argument for the equilibrated meridional temperature profile that results from baroclinic turbulence~\cite{Stone}. The concept of baroclinic adjustment originates from the analogy between baroclinic instability and upright convection: turbulent convection in the atmosphere and ocean brings the vertical profile of temperature to a state of marginal stability. This leads to a simple rule to include the effect of turbulent convection on the vertical temperature profile, called `convective adjustment': at every height, the equilibrated temperature gradient equals the critical gradient for the emergence of thermal convection. The rationale behind this result is that convection is so efficient at transporting heat that any supercritical temperature gradient will immediately be smoothed out by a large convective heat transfer that brings the system back to marginal stability. By analogy baroclinic turbulence, being a form of slantwise convection that enhances meridional heat transport, has also been argued to keep the meridional temperature profile close to marginal stability~\cite{Stone}. The `baroclinic adjustment' argument therefore predicts that the meridional temperature gradient at each latitude is given by the $\beta$-dependent critical temperature gradient. However, numerical simulations of both oceans~\cite{Jansen12} and planetary atmospheres~\cite{ZGV} suggest that the meridional temperature gradient can greatly exceed its critical value. Hydrographic sections across the Antarctic Circumpolar Current are also  characterized by supercritical temperature gradients~\cite{Tulloch,Jansen12}. The absence of zonal jets near the poles of gaseous planets, as seen in panel~\ref{fig:SO_Jupiter}a, are also inconsistent with the hypothesis that the system operates in the vicinity of the marginal stability associated with $\beta$.

Alternative arguments have thus been put forward to provide a finer description of heat transport by baroclinic turbulence through scaling predictions for the diffusivity coefficients~\cite{Held96}. However, while the insightful theoretical arguments provide qualitative understanding of the numerical data accumulated over the years~\cite{Held96,Thompson07,CH21}, they do not provide a quantitative scaling theory for the emergent diffusivity coefficients to the point that one may feel justified to question the very possibility of closing baroclinic turbulence on the basis of simple physical arguments. Building on a recent scaling theory developed for $f$-plane baroclinic turbulence~\cite{Gallet}, we turn to the crucial impact of planetary curvature and zonal jets on the $\beta$-plane, with the following motivational questions in mind:
\begin{itemize}
\item Can one express the heat flux driven by baroclinic turbulence on a curved planet using a downgradient diffusive closure? What are the key parameters the turbulent diffusivity coefficient depends on? Can we predict that dependence based on simple physical arguments? What is the role of the parameter $\beta$ associated with planetary curvature?
\item Can we leverage the emergent scale separation between the turbulent flow and the planetary radius to turn the scaling theory into a physically based parameterization of meridional heat transport by baroclinic turbulence?
\item Can the parameterization be used to quantitatively predict the meridional temperature profile in an idealized model of planetary atmosphere subject to a meridionally dependent external heat flux on a curved planet?
\item Does the resulting parameterization predict that supercritical temperature gradients can be maintained in fully turbulent planetary flows?
\end{itemize}

With the goal of answering these central questions, we turn to what is arguably the simplest model of a patch of planetary atmosphere or ocean: the two-layer QG model (2LQG in the following) on the $\beta$-plane, proposed by \citeA{Phillips} and sketched in Figure~\ref{fig:schematic}b. \cor{The atmosphere or ocean is represented by two layers of constant density, the upper layer representing more buoyant fluid sitting on top of the less buoyant lower-layer fluid. Ignoring salinity in the ocean and moisture in the atmosphere, buoyancy is proportional to potential temperature (temperature corrected for reversible compressive effects which do not contribute to heat transport). For brevity, in the following we simply refer to the upper layer as warmer than the lower layer, but the reader should keep in mind that this is to be understood in terms of potential temperature in an atmospheric context.} The two-layer system is equivalent to projecting the full dynamics of a stratified fluid onto the two gravest vertical modes which are known to contain most of the energy in fully turbulent atmospheric and oceanic flows~\cite{Flierl}. 
A drag force is imposed on the lower layer to represent drag exerted by the deep electrically conducting interior~\cite{Liu2008,Schneider2009}, or friction on bottom topographic features. In its simplest formulation, fluid columns in each layer move only horizontally under the constraint that they conserve their potential vorticity, which can be understood from the conservation of mass and angular momentum of cylindrical fluid columns in each layer as they are stretched or compressed. Coupling between the two layers comes about because any change in thickness in one layer induces an opposite change in the other layer. In the idealized picture where the upper-layer fluid is lighter because it is warmer, the height of the interface between the two layers is a direct proxy for the local vertically averaged temperature. A meridionally sloping interface is therefore equivalent to a meridional (vertically averaged) temperature gradient in 2LQG. The external differential heat flux is thus represented in 2LQG as a source/sink term that drives meridional gradients in the height of the interface. The response to this external forcing rapidly develops eddying motion through baroclinic instability, and the resulting turbulence acts to flatten the density interface. Determining the equilibrated meridional temperature profile (meridional profile of interface height) in this inhomogeneous 2LQG model is arguably the simplest situation that a parameterization of baroclinic turbulence should be able to address. And yet the nonlinear evolution results in a surprisingly peculiar instance of a turbulent flow that challenges parameterizations based on the Kolmogorov-Kraichnan spectral description of standard hydrodynamic turbulence~\cite{Salmon,Grianik}.

That a purely spectral approach may not be adequate was suggested by~\citeA{Thompson06} (TY06 in the following), who showed that the instability fuels a gas of coherent isolated vortices spanning both layers~\cite{Carnevale,Borue}. The gas of vortices reaches an equilibrium when the potential energy released to generate the vortices is balanced by frictional dissipation.
\cor{Each vortex has a very small vortical core as compared to the inter-vortex distance, and thus a broadband spectral decomposition with precise phase relations between the various spectral coefficients. This may be at the origin of the inadequacy of spectral approaches, which ignore the phase information and assume interactions restricted to neighboring wavenumbers in spectral space.
 In~\citeA{Gallet} (GF in the following), we embraced the physical-space (as opposed to the spectral-space) approach. By combining the statistics of isolated vortices and vortex dipoles with energetic arguments, we derived} a predictive scaling theory (referred to as the `vortex gas' scaling theory in the following) for the meridional heat transport as a function of external forcing and friction. However, a limitation of the study in GF is that it is restricted to the $f$-plane. It does not include planetary curvature and the associated meridional changes in the Coriolis parameter. While possibly relevant for the polar regions of gaseous planets, as illustrated in panel~\ref{fig:SO_Jupiter}a, the vortex gas theory in GF cannot describe the mid-latitude zonal jets visible in panels~\ref{fig:SO_Jupiter}b,c and would thus greatly overestimate the turbulent heat transport in those regions.

\begin{figure}[t]
\includegraphics[width=14 cm]{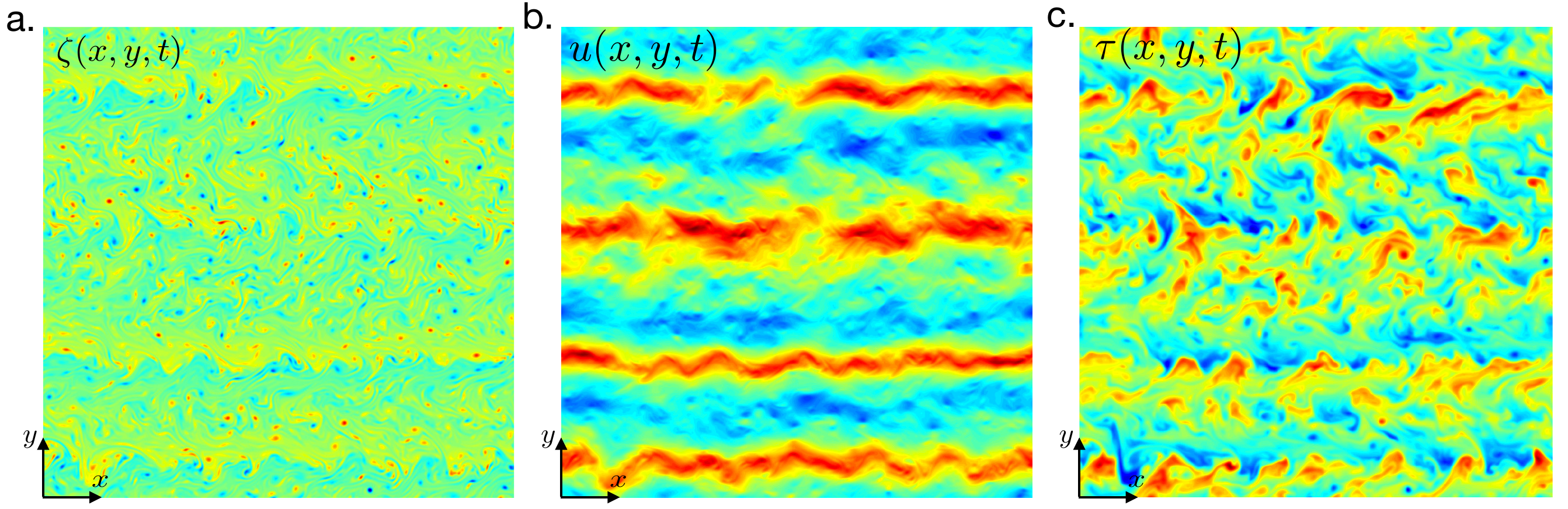} 
\caption{Snapshots of the barotropic vorticity ${\zeta = \bnabla^2 \psi}$, barotropic zonal flow ${u=-\partial_y \psi}$ and temperature $\tau$ from a numerical simulation with ${\beta_*=0.3}$ and low quadratic drag ${\mu_*=0.001}$ (fields represented in arbitrary units, low values in dark blue and large values in red). \label{fig:snapshots}}
\end{figure}

In the following, we thus augment the vortex gas scaling theory to capture the impact of planetary curvature on meridional heat transport, within the canonical 2LQG model on the $\beta$ plane. Following~\citeA{Pavan} and~\citeA{Chang}, we then show that the resulting expression for the turbulent diffusivity provides a quantitative parameterization of meridional heat transport on a rotating planet. The new theory further demonstrates that planetary atmospheres can exceed a marginally critical state, but the dependence of the criticality on the radiative forcing is very weak explaining why the baroclinic adjustment assumption gained traction. The core of the paper is devoted to the more technical aspects of the study, and the uninterested reader can jump directly to the conclusion section~\ref{sec:Conclusion}, where we discuss the main findings which answer the questions raised above.


\section{Homogeneous baroclinic turbulence on the $\beta$ plane \label{sec:Homogeneous}}

The homogeneous 2LQG model on the $\beta$-plane has been extensively described in the literature, and we only recall its salient features. The system is  sketched in Figure~\ref{fig:schematic}b. It consists in a local patch of atmosphere or ocean represented in a local Cartesian frame, the zonal, meridional and vertical directions being denoted respectively as $x$, $y$ and $z$ (see also panel~\ref{fig:schematic}a). The atmosphere or ocean is modelled as two fluid layers of equal average depth, stacked in the vertical: \cor{denser/cooler fluid in the bottom layer and lighter/warmer fluid in the upper layer (in a potential temperature sense)}. The base flow consists of vertical shear: there is a uniform zonal flow $U {\bf e}_x$ (resp. $- U {\bf e}_x$) in the upper (resp. lower) layer. Thermal-wind balance implies that the vertically sheared zonal flow must coexist with a meridional gradient of vertically averaged temperature, i.e., the interface between the two layers is tilted. On vertical average, the fluid is thus warmer toward the Southern edge of the domain than the Northern edge, see Figure~\ref{fig:schematic}b. This tilt of the interface is associated with available potential energy \cite{Lorenz} that the system releases through baroclinic instability. We consider the equilibrated state that arises from the nonlinear stage of this instability. The dynamics in each fluid layer reduces to an evolution equation for the potential vorticity (PV), a scalar field that encodes the conservation of mass and angular momentum as fluid moves on a curved rotating planet. We denote with a subscript $1$ (resp. $2$) quantities in the upper (resp. lower) layer. The departures ${\bf u}_{1;2}$ of the horizontal velocity fields (zonal and meridional components) from the base zonal flows are divergenceless at leading order and can be expressed in terms of two streamfunctions $\psi_{1;2}(x,y,t)$, i.e., ${\bf u}_{1;2}=-\bnabla \times (\psi_{1;2} \, {\bf e}_z)$. Material conservation of potential vorticity \cor{$q_{1;2}(x,y,t)$} in each layer gives the two evolution equations~\cite{Phillips,Flierl,Salmon80}: 
\begin{eqnarray}
\partial_t q_1 + U \partial_x q_1 + \left(\beta+\frac{U}{\lambda^2} \right) \partial_x \psi_1 +J(\psi_1,q_1) & = &  -\nu \bnabla^8 q_1  \label{eqq1}\\
\partial_t q_2 - U \partial_x q_2 + \left(\beta-\frac{U}{\lambda^2} \right) \partial_x \psi_2 + J(\psi_2,q_2) & = & -\nu \bnabla^8 q_2  + {\cal D} \label{eqq2} 
\end{eqnarray}
where the Jacobian is  $J(f,g)=\partial_x f \partial_y g - \partial_x g \partial_y f$. The potential vorticities are related to the streamfunctions through:
\begin{eqnarray}
q_1 & = & \boldsymbol{\nabla}^2 \psi_1 + \frac{1}{2\lambda^2}(\psi_2-\psi_1) \, , \label{defq1}\\
q_2 & = & \boldsymbol{\nabla}^2 \psi_2 + \frac{1}{2\lambda^2}(\psi_1-\psi_2) \, . \label{defq2}
\end{eqnarray}
In these expressions, $\lambda$ denotes the Rossby radius of deformation, i.e., the characteristic length scale at which rotational and temperature variation effects operate on a comparable time scale.
The dissipation terms on the right-hand side of equations (\ref{eqq1}-\ref{eqq2}) consist of hyperviscosity to dissipate filaments of potential vorticity (enstrophy) generated by eddy stirring at small scales, together with a drag term ${\cal D}$ confined to the lower-layer equation (\ref{eqq2}). As compared to standard viscosity, hyperviscosity allows one to consider large ratios between the domain size and $\lambda$, with negligible hyperviscous dissipation at the small scale $\lambda$. The drag term ${\cal D}$ corresponds to either linear or quadratic bottom friction~\cite{Thompson06,Arbic,Chang,CH21}. The former can be derived analytically as an Ekman friction term associated with Ekman pumping on a flat bottom boundary, while the latter is arguably a better model of turbulent drag on the small-scale topographic features of a rough bottom boundary. In the case of linear drag, ${\cal D} =-2 \kappa \boldsymbol{\nabla}^2 \psi_2 $, and in the case of quadratic drag, ${\cal D}=- \mu \left[ \partial_x(|\boldsymbol{\nabla} \psi_2| \partial_x \psi_2) + \partial_y(|\boldsymbol{\nabla} \psi_2| \partial_y \psi_2)  \right]$. Following previous authors, we thus neglect the cross terms between $U$ and $\psi_2$ in the quadratic drag formulation, which are shown to be negligible in GF. Alternatively, one can avoid these cross terms by considering a model where the base zonal flow is $2U {\bf e}_x$ in the upper layer and vanishes in the lower layer.

An insightful change of variables consists in introducing the barotropic and baroclinic streamfunctions, respectively $\psi=(\psi_1+\psi_2)/2$ and $\tau=(\psi_1-\psi_2)/2$. The barotropic streamfunction $\psi(x,y,t)$ is the streamfunction of the vertically averaged flow. In the QG framework, the baroclinic streamfunction $\tau(x,y,t)$ is directly proportional to the vertical displacement of the interface between the two layers. Positive $\tau$ corresponds to a locally deeper interface, hence a locally warmer depth-averaged temperature. Negative $\tau$ corresponds to a locally shallower interface, and thus a locally cooler depth-averaged temperature. In the following, we thus refer to $\tau(x,y,t)$ as the `temperature' field, keeping in mind that $\tau$ has dimensions of a streamfunction. More precisely, $\tau$ denotes the departure of the temperature field from the base state, the base zonal flow itself being associated with a base baroclinic streamfunction $-Uy$. The tilt of the interface in Figure~\ref{fig:schematic}b thus corresponds to a background meridional gradient $-U$ of vertically averaged temperature, the latter being measured in units of streamfunction.

Denoting as $\la \cdot \ra$ a 2D spatial and time average and as $\psi_x=\partial_x \psi$ the meridional barotropic velocity, our goal is to determine the meridional heat flux $\la \psi_x \tau \ra$, or equivalently the diffusivity $D=\la \psi_x \tau \ra/U$ that connects this heat flux to the background temperature gradient $-U$. The heat diffusivity $D$ is a central quantity of the homogeneous model, from which one can immediately deduce the PV diffusivities in each layer~\cite{Vallis1988}. We thus focus on $D$ in the following.

When the domain is large enough, the energy containing scale is much smaller than the domain size, which becomes irrelevant. Similarly, the heat flux is independent of the hyperviscosity $\nu$ when the latter is small enough. We thus seek $D$ as a function of the relevant dimensional parameters $U$, $\lambda$, $\beta$ and the friction coefficient $\kappa$ or $\mu$. In dimensionless from, we seek the dimensionless diffusivity $D_*=D/(U\lambda)$ as a function of the dimensionless beta coefficient, $\beta_*=\beta \lambda^2 /U$, and the dimensionless friction coefficient, $\kappa_*= \kappa \lambda/U$ or $\mu_*=\mu \lambda$ (for linear and quadratic drag, respectively).
A related quantity of interest is the mixing length $\ell=\sqrt{\la \tau^2 \ra}/U$. This is the typical distance travelled by a fluid element carrying its background temperature, before it is mixed with the environment and relaxes to the local background temperature. It follows that the typical temperature fluctuations around the background gradient are of the order of $U \ell$. The dimensionless mixing length $\ell_*=\ell/\lambda$ is again a function of $\beta_*$ and $\kappa_*$ or $\mu_*$, depending on the form of the drag.

In GF, we characterized the dependence of the turbulent diffusivity and mixing length on the dimensionless drag coefficient, focusing on the situation $\beta=0$. We developed a scaling theory for the low-drag regime that agrees quantitatively with numerical results of the homogeneous model. 
In the case of linear drag, the scaling behaviors of the diffusivity and mixing length are: 
\begin{eqnarray}
D_* & = & 2.0 \, e^{0.72/\kappa_*} \, , \label{Dlinnobeta}\\
\ell_* & = & 2.5 \, e^{0.36/\kappa_*} \, . \label{elllinnobeta}
\end{eqnarray}
For quadratic drag we obtained:
\begin{eqnarray}
D_* & = & \frac{2.0}{\mu_*} \, , \label{Dquadnobeta}\\
\ell_* & = & \frac{2.5}{\sqrt{\mu_*}} \, . \label{ellquadnobeta}
\end{eqnarray}
The dimensionless prefactors provided in (\ref{Dlinnobeta}-\ref{ellquadnobeta}) differ by approximately $20\%$ from the ones provided in GF. The present prefactors provide a slightly better agreement to the low-drag data in GF, for which the scaling theory has been developed, whereas the values given in GF provide a somewhat better overall agreement when moderately low values of the drag are included (the differences in prefactors having a negligible impact on what follows).
We then showed in GF that the scaling-laws above could be used as a quantitative parameterization in an inhomogeneous model driven by an imposed meridionally dependent heat flux. However, any scaling theory to be used as a parameterization of mid-latitude meridional heat transport should include the key parameter $\beta$, which motivates the present study. More precisely, we address the crucial role of $\beta$ through the following questions: starting from the situation $\beta_*=0$, how large need $\beta_*$ be to impact meridional heat transport? How much lower is the resulting diffusivity, and can we augment the \cor{vortex gas} scaling theory to incorporate the crucial parameter $\beta_*$? Finally, is the resulting scaling theory a skillful parameterization for the inhomogeneous two-layer QG model with meridionally dependent large-scale forcing?

\begin{figure}[t]
\hspace{-1.5 cm}
\includegraphics[width=8 cm]{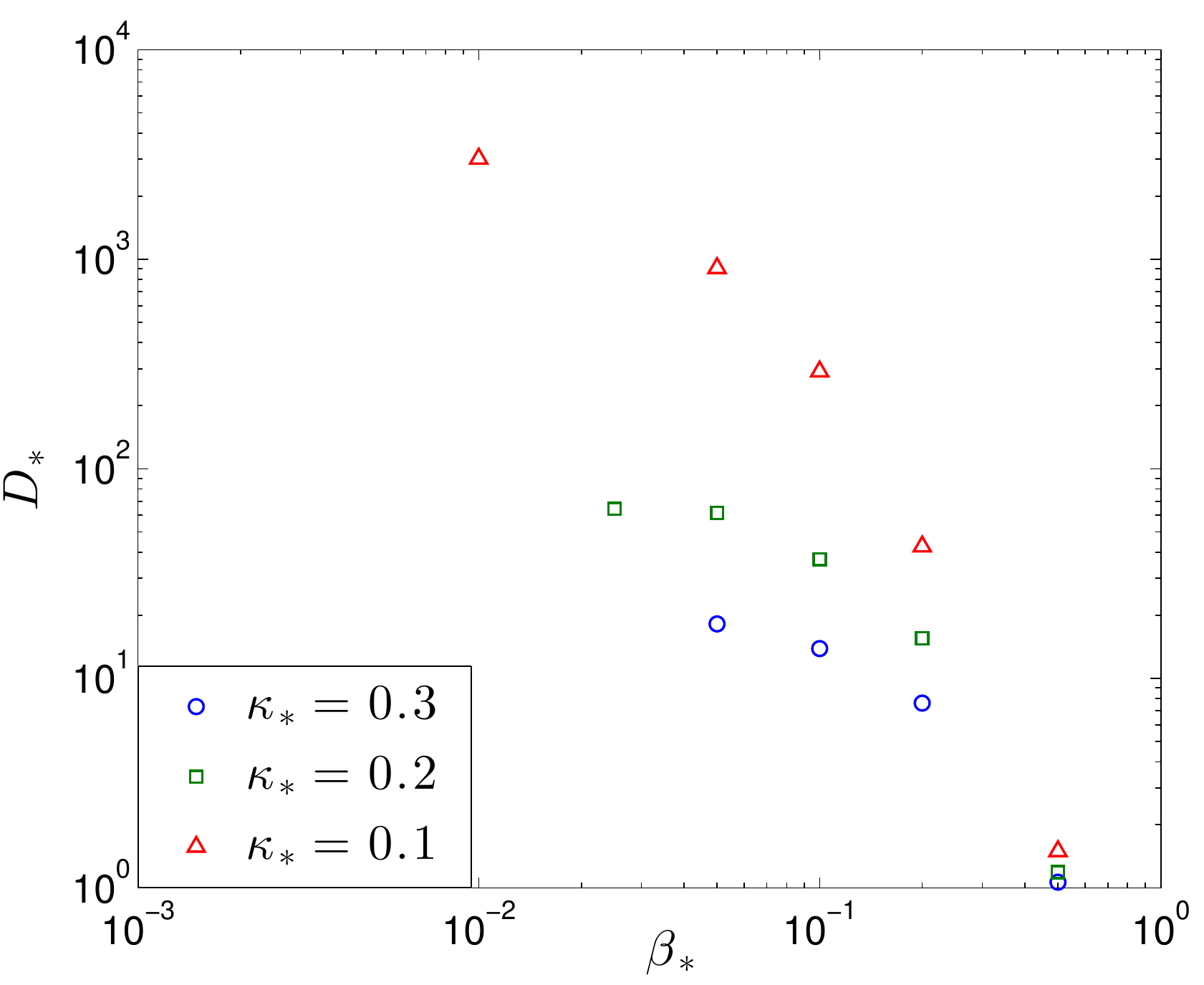} 
\includegraphics[width=8 cm]{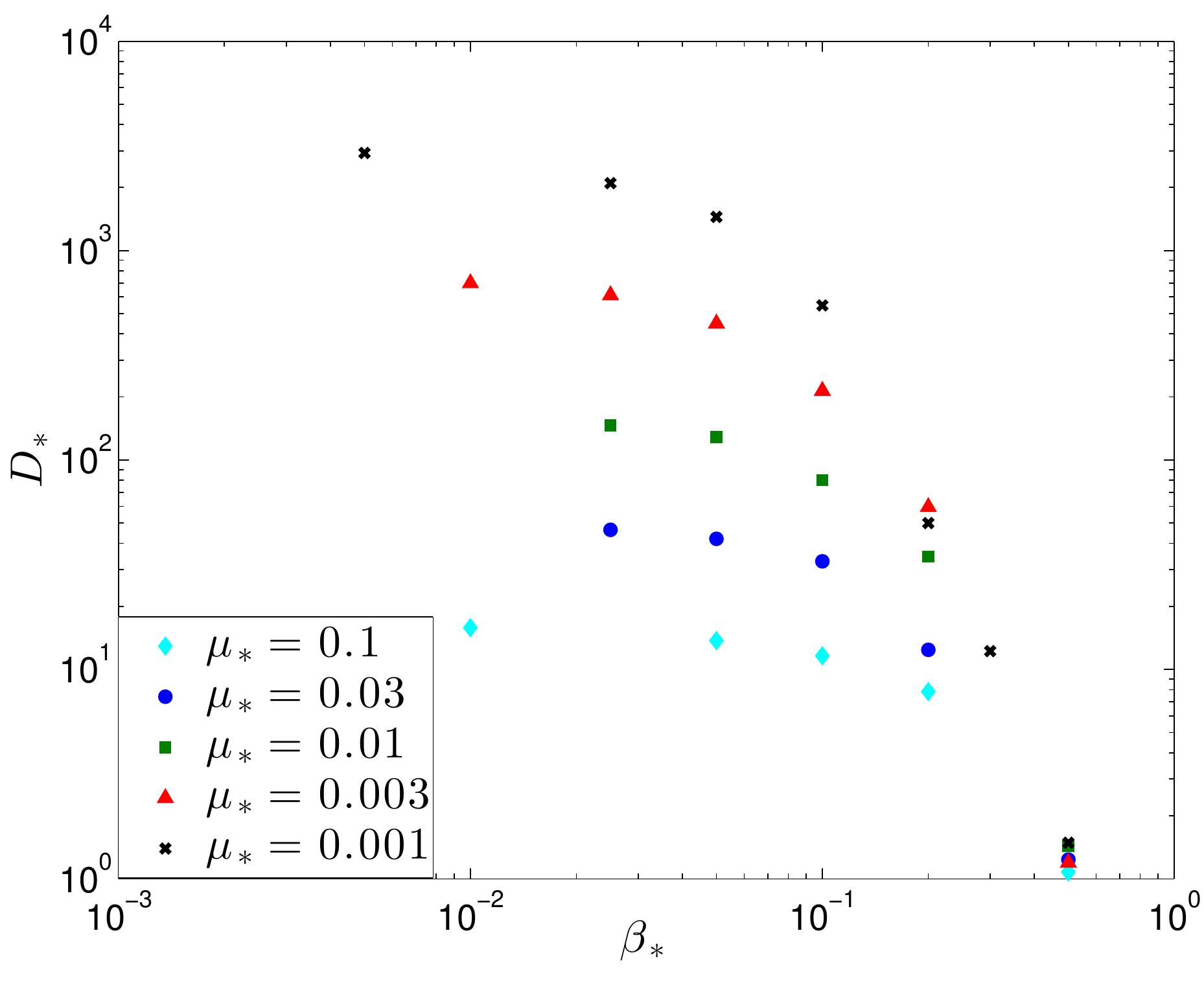} 
\caption{Dimensionless diffusivity versus dimensionless $\beta$ for linear drag (left) and quadratic drag (right). In both cases $D_*$ decreases rapidly with increasing $\beta_*$ to reach values of the order of one for ${\beta_* \lesssim 1}$.   \label{fig:D_vs_beta}}
\end{figure}

\section{Suppression of meridional transport by $\beta$}

To address these questions, we performed a suite of numerical simulations of equations (\ref{eqq1}-\ref{eqq2}) inside a large domain with weak hyperviscosity. 
Because we aim at modelling a local patch of atmosphere or ocean, and because this patch is much larger than the energy containing scale, one can conveniently use periodic boundary conditions in the horizontal directions for the departure fields. The details of the numerical procedure are given in \ref{app:Num}. We extract the diffusivity $D_*$ in statistically steady state, for several values of the -- linear or quadratic -- drag coefficient, and $\beta_*$. We focus on the strongly turbulent regime that arises for values of $\beta_*$ significantly lower than the threshold value $\beta_*=1$ at which the undamped system is no longer unstable. As $\beta_*$ approaches one, the diffusivity $D_*$ drops to values of order one: this is the weakly nonlinear regime, where the precise parametric dependence of $D_*$ becomes more complex as documented by~\citeA{Thompson07} and~\citeA{CH21}. The present suite of numerical simulations thus focuses on $\beta_* \in [0;0.5]$, which corresponds to the strongly turbulent, large-diffusivity and large-mixing-length regime for which we derived the vortex gas scaling theory in GF. The only aspect of the weakly nonlinear regime that will enter the development of the scaling theory is the expectation that $D_*$ decreases to values ${\cal O}(1)$ as $\beta_*$ approaches one, irrespective of the value of the drag coefficient.


The diffusivity $D_*$ is shown as a function of $\beta_*$ in Figure \ref{fig:D_vs_beta}, for several values of the drag coefficient. The qualitative behavior is the same for linear and quadratic drag: $\beta_*$ drastically suppresses meridional heat transport, with a diffusivity reduced by orders of magnitude as $\beta_*$ increases. The lower the drag, the higher the $\beta_*=0$ initial value of the diffusivity (in line with equations (\ref{Dlinnobeta}) and (\ref{Dquadnobeta})), and the lower the values of $\beta_*$ that affect meridional transport. As $\beta_*$ increases, the flow evolves from the $\beta_*=0$ barotropic vortex gas to a combination of isolated vortices and coherent zonal jets characteristic of $\beta$-plane turbulence. We illustrate this coexistence of vortices and jets in Figure~\ref{fig:snapshots}, which displays snapshots of barotropic vorticity, barotropic zonal flow and temperature for $\beta_*=0.3$ and quadratic drag $\mu_*=0.001$.

A second key observation is that, on the logarithmic plot of Figure \ref{fig:D_vs_beta}, the curves corresponding to various values of the drag seem to converge to values of order one when $\beta_*$ itself becomes ${\cal O}(1)$. In other words, as $\beta_*$ increases, the diffusivity drops from the large values of the vortex gas scaling regime to the ${\cal O}(1)$ values characteristic of the weakly nonlinear regime that arises in the vicinity of the $\beta_*=1$ frictionless threshold for baroclinic instability (see also \citeA{Thompson07,CH21}).


\section{Frictional vs. zonostrophic arrest of the inverse energy cascade}

Our goal is to augment the \cor{vortex gas} scaling theory to capture the suppression of meridional transport by $\beta$. The first step consists in seeking the control parameter that governs the impact of $\beta$ on meridional heat transport. In particular we will begin by determining for what value $\beta_*>0$ the meridional heat transport starts decreasing below its value for $\beta_*=0$. Conveniently, we can use the vortex gas scaling theory to answer that question, because by definition it remains valid up to the transition value of $\beta_*$ at which $\beta$ begins to affect the dynamics.

As a preamble, we briefly recall the main scaling relations of the $\beta_*=0$ vortex gas, with the goal of determining the peak wavenumber $k_0$ of the barotropic energy spectrum. As described in TY06 and GF, the barotropic flow for $\beta_*=0$ consists in a gas of isolated vortices. The vortex core radius is close to $\lambda$, the inter-vortex distance scales with the mixing length $\ell \gg \lambda$, and the typical circulation $\pm \Gamma$ of the vortices scales with the diffusivity $D$, which itself scales with the squared mixing length: $\Gamma/(U \lambda) \sim D_* \sim \ell_*^2$. The mean square velocity of such a vortex gas is given by $\la {\bf u}^2 \ra \sim (D/\ell)^2 \log \ell_*  \sim U^2 \ell_*^2 \log \ell_*$, where ${\bf u}$ denotes the barotropic velocity field and the logarithmic term stems from the large velocities arising at the periphery of the vortex core. The mean-squared streamfunction can be estimated by focusing on a single dipole of vortices of circulations $\pm \Gamma$ and separated by a distance $\ell$. One obtains simply $\la \psi^2 \ra \sim \Gamma^2 \sim D^2$.

In spectral space, TY06 convincingly show that the energy-containing wavenumber $k_0$ where the barotropic energy spectrum peaks can be inferred from the ratio of the barotropic kinetic energy to the mean squared barotropic streamfunction: $k_0=\sqrt{\la {\bf u}^2 \ra/\la \psi^2 \ra}$. Using the vortex gas estimates above, we obtain:
\begin{eqnarray}
k_0 \sim \frac{\log^{1/2} \ell_*}{\ell} \, .
\end{eqnarray}
The logarithmic factor at the numerator is at the origin of TY06's claim that the energy containing wavenumber $k_0$ differs from the inverse mixing length, challenging an earlier statement by~\citeA{Larichev} (after inserting expression (\ref{elllinnobeta}) for $\ell_*$, the vortex-gas prediction $k_0 \ell \sim \log^{1/2} \ell_*$ captures very well the data for $k_0 \ell$ provided in Figure~6a of TY06). In the $\beta_*=0$ vortex gas regime, the barotropic inverse energy transfers tend to increase the inter-vortex distance $\ell$, thus reducing the energy containing wavenumber $k_0$. Such an inverse energy cascade is arrested by bottom drag, which ultimately sets the equilibrated values of $\ell$ and $k_0$. This is the `frictional' arrest mechanism~\cite{Grianik,Borue,TsangYoung,Tsang}.

In contrast to this situation, $\beta$-plane turbulence offers a competing arrest mechanism for the inverse energy cascade~(\citeA{Held96}, HL in the following): as described at the outset, $\beta$ tends to channel the inverse energy flux into zonal jets, i.e., intense flow structures that do not contribute to meridional transport~\cite{Maltrud}. As initially shown by~\citeA{Rhines75}, this `zonostrophic' arrest mechanism yields a barotropic energy spectrum that peaks at the Rhines wavenumber $k_R=\sqrt{\beta/\la {\bf u }^2\ra^{1/2}}$.

We are thus in a position to determine the control parameter that governs the reduction of meridional transport by $\beta$: for low $\beta_*$, the Rhines wavenumber $k_R$ is much lower than $k_0$, and the frictional arrest mechanism operates before the flow even feels the presence of $\beta$. However, as $\beta_*$ increases, so does $k_R$, until the latter becomes comparable to $k_0$. As $\beta_*$ further increases, the zonostrophic mechanism takes over; we expect the emergence of Rossby waves and jets together with reduced meridional transport. As we increase $\beta_*$ from zero, which arrest mechanism dominates is thus governed by the ratio of $k_R$ and $k_0$, or equivalently by the ratio $(k_R/k_0)^2$. This line of argument is similar to the one in HL and \citeA{CH21}, one crucial difference being that $k_0$ is not being evaluated within a spectral framework but rather within the framework of the vortex gas model, which holds from $\beta_*=0$ up to the transition point where $\beta$ starts affecting the dynamics. We thus estimate the ratio $k_R^2/k_0^2$ by substituting the $\beta_*=0$ vortex-gas scaling-law for $\la {\bf u}^2 \ra$ and $\ell_*$. This leads to the following definition for the nondimensional control parameter $B$ (proportional to $k_R^2/k_0^2$):
\begin{eqnarray}
B  \equiv  \frac{\beta_* \lb0}{\log^{3/2} \lb0} \, , \label{Bgeneral}
\end{eqnarray}
\cor{where $\lb0$ denotes the value of $\ell_*$ for $\beta=0$, given by (\ref{elllinnobeta}) or (\ref{ellquadnobeta}) depending on the form of the drag (the subscript `vg' referring to the $\beta=0$ vortex gas theory)}. The control parameter $B$ thus depends only on $\beta_*$ and the friction coefficient through $\lb0$. For linear and quadratic drag respectively, it takes the form:
\begin{eqnarray}
B = \left\{ \begin{matrix}
 \frac{2.5 \, e^{0.36/\kappa_*} \beta_*}{ \log^{3/2} \left(2.5 \, e^{0.36/\kappa_*} \right)} \qquad \text{(linear drag)} \\
 \\
\frac{2.5 \, \beta_*}{\sqrt{\mu_*} \log^{3/2}\left( \frac{2.5}{\sqrt{\mu_*}} \right)}  \qquad \text{(quadratic drag)}
\end{matrix} \right. \label{eq:commonB}
\end{eqnarray}
To summarize, we have identified the control parameter $B$ based on a competition between the frictional and zonostrophic arrest mechanisms for the inverse energy cascade. As $\beta_*$ increases, we expect a reduction of the diffusivity when $B$ becomes of order one (in the sense commonly assumed in scaling analysis, that is we expect the transition value of $B$ to be between $0.1$ and $10$). Strictly speaking, we have only shown that $B$ is the right control parameter up to the transition point where $\beta$ starts impacting the dynamics. Beyond that transition, jets and Rossby waves emerge and one cannot {\it a priori} use the $\beta=0$ vortex-gas scaling laws. Nevertheless, we will show in the next section that $B$ is also the right control parameter in the large-$\beta$ regime.

Another interpretation for the parameter $B$ is as the inverse Rossby wave steepness $(k_0/k_R)^2$~\cite{Rhines75}, a measure of the nonlinearity of the Rossby waves excited by the vortex gas. For low $B$, these waves are extremely nonlinear. They are distorted by the vortex gas within less than a wave period, which means that wave dynamics is hardly detectable. By contrast, large $B$ allows for wave dynamics and the associated wave-wave interactions that induce zonal jets and reduced meridional velocity.

\begin{figure}[]
\hspace{-2.3 cm}
\includegraphics[width=18 cm]{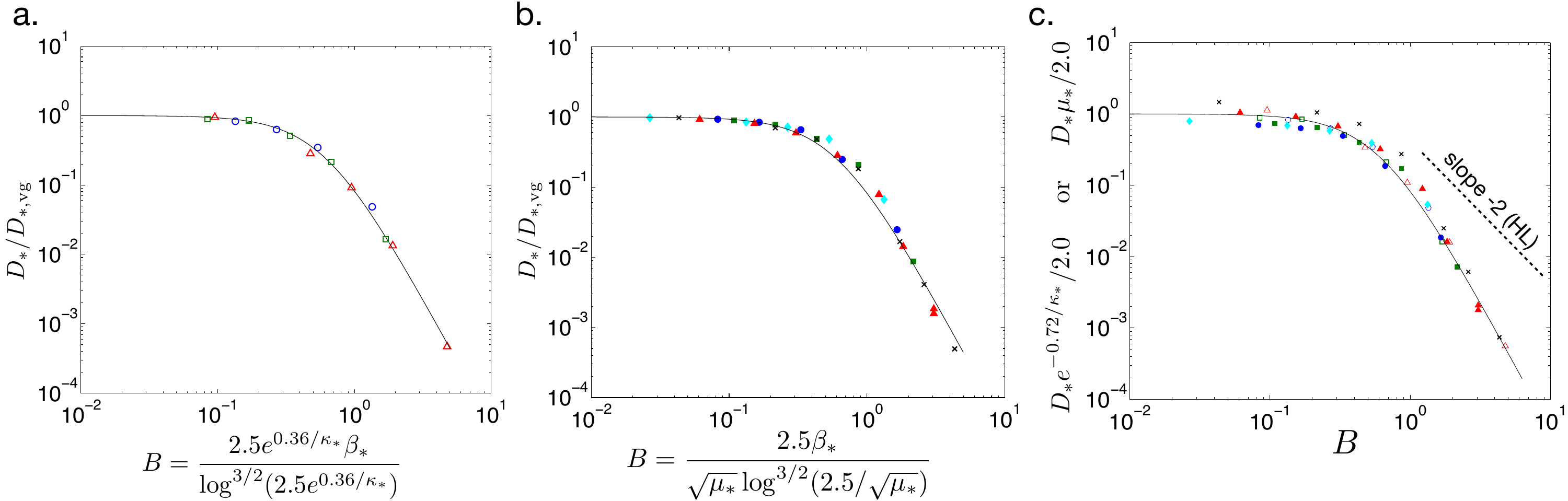} 
\caption{Dimensionless diffusivity rescaled by its ${\beta_*=0}$ value, $D_*/\Db0$, as a function of the control parameter $B$. \cor{In panels a and b, $\Db0$ is obtained from a numerical simulation with ${\beta_*=0}$.} This representation leads to a collapse of the data for both linear drag (a) and quadratic drag (b). \cor{In panel c, we replace $\Db0$ by its expression in the GF vortex gas theory,} either (\ref{Dlinnobeta}) or (\ref{Dquadnobeta}) depending on the type of drag. The data for both linear and quadratic drag are represented on this panel. They collapse onto a single master curve in this representation. The solid line shows the theoretical expression (\ref{fullF}), which captures the master curve in all three panels, while the dashed-line in panel c indicates the scaling exponent predicted by HL \cor{(they predict a dimensionless diffusivity decreasing as $\beta_*^{-2}$ irrespective of the drag)}. The definitions of $B$ for linear and quadratic drag are recalled in the $x$-labels of panels a and b, and the symbols are the same as in Figure~\ref{fig:D_vs_beta}. \label{fig:rescaledD}}
\end{figure}

To check that $B$ is indeed the control parameter that governs the reduction of meridional transport by $\beta$, we plot in Figure~\ref{fig:rescaledD} the diffusivity $D_*$, rescaled by its value $\Db0$ in the absence of $\beta$, as a function of the control parameter $B$. This representation leads to an excellent collapse of the data onto a single master curve, for both linear and quadratic drag. As $B$ increases, the master curve starts decreasing when $B$ becomes ${\cal O}(1)$. These are clear indications that we have identified the right control parameter, for both linear and quadratic drag. 
The suppression of meridional transport by $\beta$ is thus captured by the scaling relation:
\begin{eqnarray}
D_* & = & \Db0 \times {\cal F}(B) \label{defF} \\
& \sim & \lb0^2 \, {\cal F} \left(  \frac{\beta_* \lb0}{\log^{3/2} \lb0} \right) \, , \label{Dvsl0}
\end{eqnarray}
where we have substituted the scaling relation $\Db0\sim \lb0^2$ to obtain the second line. While our theoretical argument has so far demonstrated that $B$ is the right control parameter up to the transition point $B \sim 1$, the numerical data further indicate that $B$ is the right control parameter also beyond the transition point. In the next section we provide theoretical arguments confirming that $B$ is the right control parameter throughout the entire parameter space considered in this study, before determining the master function ${\cal F}$.

\section{The drag-independent zonostrophic regime}

\begin{figure}[]
\centering
\includegraphics[width=10 cm]{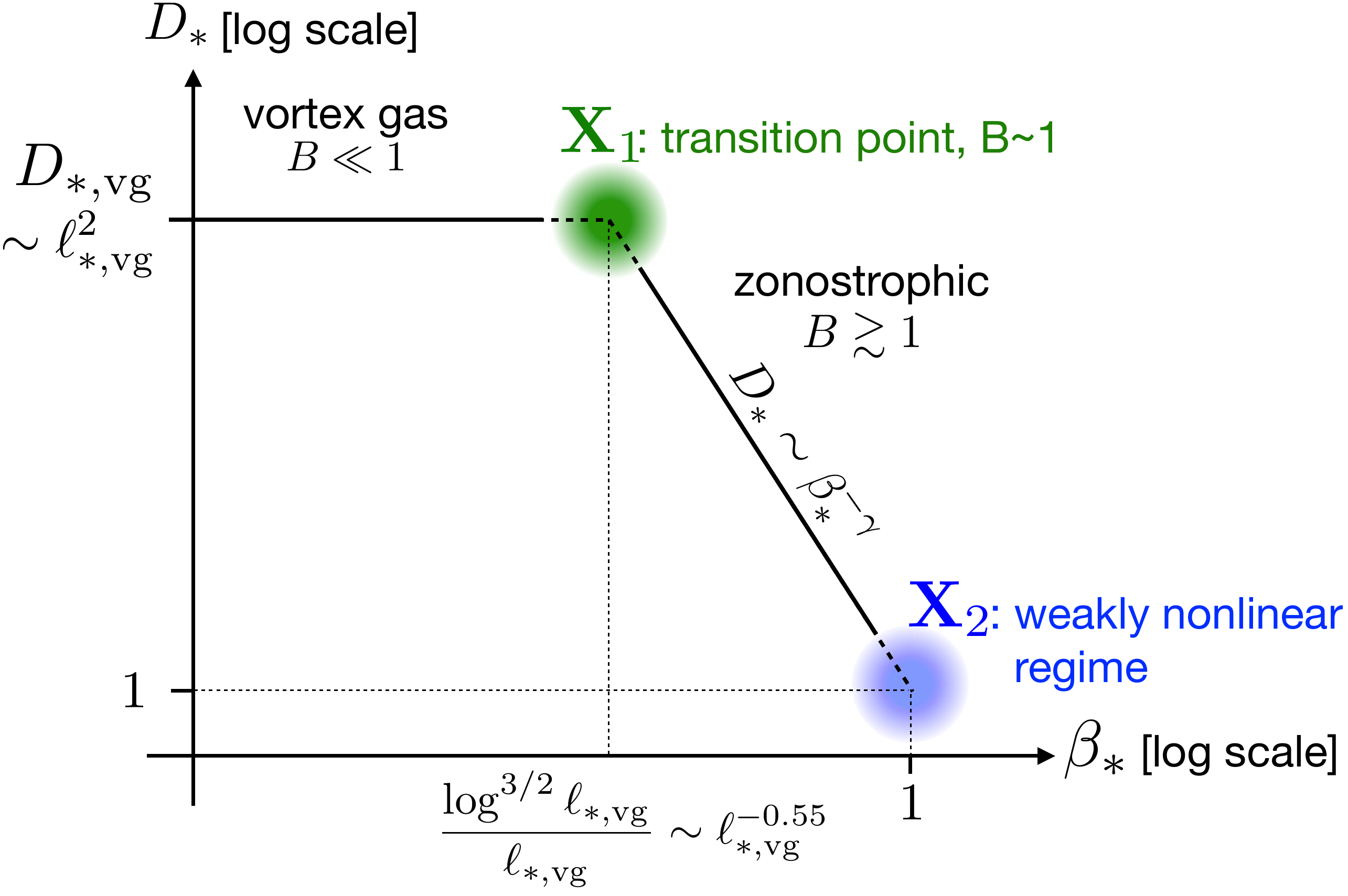} 
\caption{Schematic of the two scaling regimes arising in Figure~\ref{fig:D_vs_beta}. The vortex-gas scaling regime holds up to the crossover point ${\bf X}_1$, characterized by ${B \sim 1}$. Beyond this point, the zonostrophic scaling regime connects ${\bf X}_1$ to the weakly nonlinear regime ${\bf X}_2$, characterized by ${D_*\sim 1}$ and ${\beta_* \sim 1}$. Because we know the coordinates of ${\bf X}_1$ and ${\bf X}_2$, we can determine the scaling exponent $\gamma$ of the zonostrophic scaling regime. The actual crossover at ${\bf X}_1$ is smoother than represented here.\label{fig:schematic_scaling}}
\end{figure}

As $B$ reaches values of order one, the system enters the `zonostrophic' regime. This regime connects the $\beta_*=0$ vortex-gas regime to the weakly nonlinear regime that arises for $\beta_*={\cal O}(1)$. In the latter regime, $D_*$ takes values of order one, regardless of the (weak) drag coefficient. As a matter of fact, \citeA{Thompson07}(TY07 in the following) identify a `pivot point' $\beta_*=\beta_p\simeq 0.6$ in their numerical data where the diffusivity  $D_*=D_c={\cal O}(1)$ is independent of the linear drag coefficient (and after which $D_*$ becomes an increasing function of the drag coefficient). The pivot point corresponds approximately to the rightmost data point in Figure~\ref{fig:D_vs_beta}, for both linear and quadratic drag. The very existence of the pivot point may be specific to the 2LQG model, but the physical insight that $D_*$ becomes ${\cal O}(1)$ as $\beta_*$ approaches ${\cal O}(1)$ values is expected to be a robust feature across various models.

In Figure~\ref{fig:schematic_scaling}, the vortex-gas and zonostrophic regimes are represented schematically in the $(\beta_*,D_*)$ plane. As $\beta_*$ increases, the vortex gas regime holds up to the transition point ${\bf X}_1$ characterized by $B\sim 1$ and $D_*\sim \Db0$. Its coordinates are thus  \ $(\beta_* \sim \lb0^{-1} \log^{3/2} \lb0 \, , \quad D_* \sim \lb0^2)$ in the $(\beta_*,D_*)$ plane. Beyond this transition point, i.e., for $B\gtrsim 1$, the zonostrophic regime connects ${\bf X}_1$ to the weakly nonlinear regime characterized by the point ${\bf X}_2$ of approximate coordinates $(\beta_* \sim 1, D_*\sim 1)$. For $B\gtrsim 1$ we expect the zonostrophic arrest mechanism to be fully operational and the drag coefficient to become irrelevant (this idea is at the core of the theory of HL and \citeA{CH21}). The scaling between ${\bf X}_1$ and ${\bf X}_2$ indeed appears to follow a drag-independent power-law $D_* \sim \beta_*^{-\gamma}$, the exponent of which we now predict. Following standard practice in turbulence research, the logarithmic factor in the coordinates of ${\bf X}_1$ can be approximated as a (significant) correction to the local exponent of a power-law behavior over the whole $\lb0 \in [8 ; 100]$ range spun in our simulations:
\begin{eqnarray}
 \frac{\lb0}{\log^{3/2} \lb0} \simeq 0.80 \,  \lb0^{\alpha} \, , \label{approxlog}
\end{eqnarray}
which holds within $10\%$, with $\alpha=0.55$. The coordinates of the transition point ${\bf X}_1$ are then approximated by $(\beta_*\sim \lb0^{-\alpha}, \quad D_*\sim \lb0^2)$. Thus at the transition point $D_*\sim \beta_*^{-2/\alpha}$ and this relationship also holds at ${\bf X}_2$ where $D_*\sim\beta_*\sim 1$. If a power-law is to hold at ${\bf X}_1$, ${\bf X}_2$ and in between it must therefore be of the form $D_* \sim \beta_*^{-\gamma}$ with exponent $\gamma=2/\alpha\simeq3.64$. Additionally, using (\ref{approxlog}) we can write:
\begin{eqnarray}
\Db0  \sim \lb0^2 \sim \lb0^{\alpha \gamma} \sim \left(  \frac{ \lb0}{\log^{3/2} \lb0} \right)^\gamma \, .
\end{eqnarray}
Dividing the scaling relation $D_* \sim \beta_*^{-\gamma}$ by the equality above leads to:
\begin{eqnarray}
\frac{D_*}{\Db0} \sim B^{-\gamma} \, . \label{asymptF}
\end{eqnarray}
We conclude that $B$ is the control parameter also in the zonostrophic regime. This provides {\it a posteriori} justification that formulating the scaling theory with the simple form (\ref{defF}) does not require an assumption that the vortex gas scaling-laws hold in the zonostrophic regime $B \gtrsim 1$.

To summarize, we have first identified the control parameter $B$ by demanding that $k_R$ and $k_0$ be comparable when the effect of $\beta$ comes into play. We have used the vortex-gas scaling theory to determine $B$, because it remains valid in the frictional regime between $\beta=0$ and the transition point $B \sim 1$. We then derived the zonostrophic scaling-law by demanding that the transition point be connected to the weakly nonlinear regime achieved for $\beta_* \sim 1$ in a way that does not involve the drag coefficient. This leads to the power-law $D_* \sim \beta_*^{-\gamma}$ with $\gamma \simeq 3.64$, which can also be written under the form (\ref{asymptF}). We thus conclude that ${D_*}/{\Db0}$ is a function ${\cal F}(B)$ over the entire region of parameter space considered in this study, with ${\cal F}(0)=1$ and ${\cal F}(B)\sim B^{-\gamma}$ for $B \gtrsim 1$.


\begin{figure}[t]
\hspace{-1.5 cm}
\includegraphics[width=17 cm]{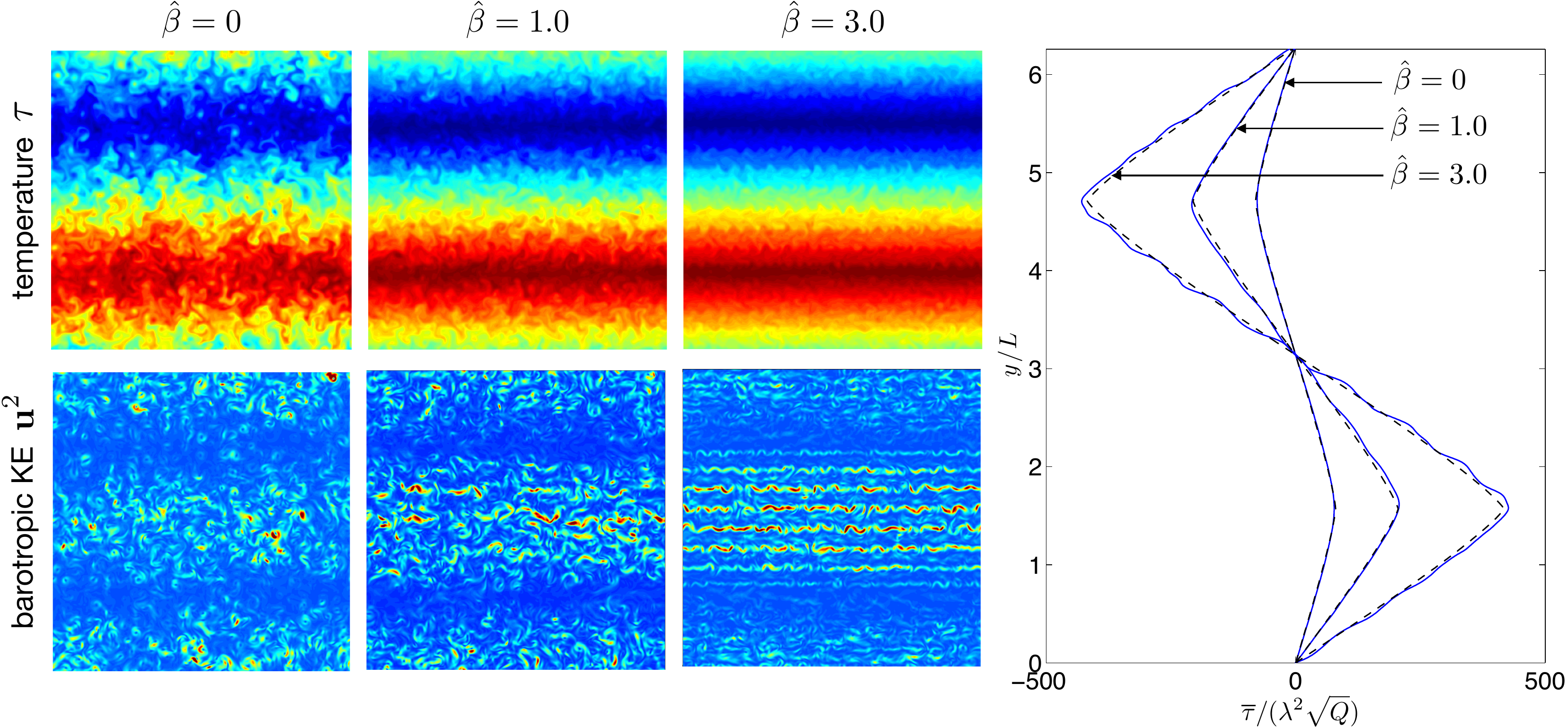} 
\caption{Snapshots of the temperature and barotropic kinetic energy in the equilibrated state of the inhomogeneous model, for increasing values of ${\hat{\beta}=\beta \lambda / \sqrt{Q}}$ with linear drag (${\hat{\kappa}=\kappa/\sqrt{Q}=0.3}$ and ${\lambda/L=0.02}$, colorplots in arbitrary units with large values in red and low values in blue). The zonally averaged temperature profiles (blue solid lines) are accurately captured by the GF profile for ${\hat{\beta}=0}$, and by the theoretical profile (\ref{tautruegamma}) for nonzero $\hat{\beta}$ (dashed lines). \label{fig:profiles}}
\end{figure}

The overall shape of the function ${\cal F}$ can finally be obtained through a simple matching of the vortex gas regime ${\cal F}(B)=1$ for low $B$ to the asymptotic regime (\ref{asymptF}) for large $B$. While there are many possible choices to perform this matching, the simple expression:
\begin{eqnarray}
{\cal F}(B)=\frac{1}{(1+2.5\, B^{\gamma/2})^{2}} \, , \label{fullF}
\end{eqnarray}
provides a good trade-off between simplicity and accuracy, the \cor{constant} $2.5$ at the denominator being an adjustable parameter. As shown in Figure \ref{fig:rescaledD}, adjusting this single coefficient allows us to capture the dependence of the diffusivity on $\beta$ over the entire dataset, which spans two orders of magnitude in $D_*$. That the same function ${\cal F}(B)$ correctly captures the behavior of $D_*$ for both linear and quadratic drag gives us additional confidence in the present scaling theory. Inserting the functional form (\ref{fullF}) into (\ref{defF}), using the approximation (\ref{approxlog}) for the logarithmic term in $B$, yields the following compact expressions for the diffusivity:
\begin{eqnarray}
D_* & = & \frac{2.0 }{ \left[e^{-0.36/\kappa_*} +4.2 \, \beta_*^{\gamma/2}   \right]^2} \, , \label{Dsimplelin} \\
D_* & = & \frac{2.0 }{ \left[\sqrt{\mu_*} +4.2 \, \beta_*^{\gamma/2}   \right]^2} \, , \label{Dsimplequad}
\end{eqnarray}
for linear and quadratic drag, respectively. The same approach can be used to determine the scaling behavior of the mixing length $\ell_*$, as shown in \ref{app:Red}.

\section{The inhomogeneous model}

We would like to demonstrate the skill of the scaling-laws (\ref{Dsimplelin}-\ref{Dsimplequad}) used as diffusive parameterizations when a forcing mechanism drives large-scale temperature inhomogeneities. We thus consider the 2LQG system subject to a latitudinally dependent imposed heat flux:
\begin{eqnarray}
\partial_t q_1 + J(\psi_1,q_1) + \beta \partial_x \psi_1 & = &  Q \sin(y/L) -\nu \bnabla^8 q_1 \, , \label{q1flux}\\
\partial_t q_2 + J(\psi_2,q_2) + \beta \partial_x \psi_2 & = & -Q \sin(y/L) -\nu \bnabla^8 q_2  + {\cal D} \label{q2flux} 
\end{eqnarray}
where the PVs are still related to the streamfunctions through (\ref{defq1}-\ref{defq2}). Inhomogeneous models of this kind were initially introduced by \citeA{Pavan} (see also~\citeA{Chang}). The set of equation (\ref{q1flux}-\ref{q2flux}) is the $\beta\neq0$ extension of a similar inhomogeneous model introduced in GF. In contrast to the homogeneous model, there is no base flow in the inhomogeneous model: the PVs $q_{1}$, $q_2$ and the streamfunctions $\psi_1$, $\psi_2$, $\psi$ and $\tau$ now refer to the total fields (as opposed to departures from a base state). When forming the sum and difference of (\ref{q1flux}) and (\ref{q2flux}), the $Q$ terms cancel out from the $\psi$ evolution equation, while a meridionally dependent source term arises in the $\tau$ equation. This source term  mimics the differential heat flux forcing a turbulent atmosphere.

In Figure~\ref{fig:profiles}, we show snapshots of the temperature and barotropic kinetic energy in the equilibrated state of a numerical integration of (\ref{q1flux}) and (\ref{q2flux}), for increasing values of ${\beta}$. For nonzero ${\beta}$, we notice again the emergence of zonal jets, together with a reduction in mixing length, noticeable through the smaller characteristic scale of the turbulent temperature fluctuations. From such simulations, one can extract the meridional temperature profile $\overline{\tau}(y)$, where the overbar denotes a zonal and time average. This temperature profile, as well as the vertically sheared zonal flow $U(y)=-\partial_y \overline{\tau}$, are emergent quantities in the inhomogeneous model. The goal of a parameterization of turbulent transport is to predict the strength and structure of $\overline{\tau}(y)$ as  a function of $\beta$, the strength $Q$ of the imposed heat flux and the drag coefficient. In dimensionless form, we seek the dimensionless temperature profile $\overline{\tau}(y)/(\lambda^2 \sqrt{Q})$ in terms of $y/L$, $\hat{\beta}=\beta \lambda / \sqrt{Q}$ and $\hat{\kappa}=\kappa/\sqrt{Q}$ or $\mu_*=\mu \lambda$.

Averaging the layer equations, neglecting the dissipative terms, yields:
\begin{eqnarray}
\partial_y \overline{\psi_x \tau} = - Q \lambda^2 \sin(y/L) \, . \label{avgeq}
\end{eqnarray}
A remarkable aspect of the snapshots in Figure~\ref{fig:profiles} is that a large scale separation is maintained in the turbulent state of this system: the mixing length is much less than $L$, and we thus expect the homogeneous model to hold locally. The local meridional heat flux is thus related to the local meridional temperature gradient through the diffusivity $D_*$:
\begin{eqnarray}
\overline{\psi_x \tau} = - D \partial_y \overline{\tau} = - D_* \lambda |\partial_y \overline{\tau}| \partial_y \overline{\tau} \, .
\end{eqnarray}
Focusing on a region where $\partial_y \overline{\tau}<0$ to alleviate notations, we integrate equation (\ref{avgeq}) into:
\begin{eqnarray}
D_*  (\partial_y \overline{\tau})^2 = Q \lambda L \cos(y/L) \, . \label{fluxadim}
\end{eqnarray}
The goal here is to demonstrate the skill of the augmented scaling theory to predict the strength and structure of the emerging temperature profile when $\beta \neq 0$ (see GF for the $\beta=0$ limit). We thus focus on the finite-$\beta_*$ low-drag zonostrophic regime where the drag coefficient becomes irrelevant. In this regime, both scaling-laws (\ref{Dsimplelin}) and (\ref{Dsimplequad}) reduce to the power-law behavior $D_*=c_\gamma \beta_*^{-\gamma}$, with $c_\gamma = 0.113$ \cor{($c_\gamma$ is directly related to the coefficient $2.5$ in expression (\ref{fullF}) and thus is not a new fitting parameter. See the conclusion section for a summary of the fitting parameters arising in the augmented scaling theory)}.
Substitution of this power-law into (\ref{fluxadim}), with $\beta_*=\beta \lambda^2/\partial_y \overline{\tau}$, yields:
\begin{eqnarray}
|\partial_y \overline{\tau} | =\left( \frac{Q L \lambda^{2\gamma+1} \beta^\gamma}{c_\gamma} \right)^{\frac{1}{2+\gamma}} \cos^{\frac{1}{2+\gamma}}(y/L) \, . \label{eqdytaugamma}
\end{eqnarray}

An integration over $y$ leads to the theoretical prediction for the temperature profile, given by equation (\ref{tautruegamma}) in~\ref{app:Eme}. This theoretical profile is compared to the ones obtained numerically in Figure \ref{fig:profiles}. We focus on linear drag with $\hat{\kappa}=\kappa/\sqrt{Q}=0.3$, $\lambda/L=0.02$, and increasing values of $\hat{\beta}$: $\hat{\beta}=0$, $\hat{\beta}=1$ and $\hat{\beta}=3$. For $\hat{\beta}=0$, the resulting numerical profile is compared to the $\beta=0$ theoretical prediction in GF. For $\hat{\beta}=1$ and $\hat{\beta}=3$, we compare the numerical profile to (\ref{tautruegamma}). In all cases the agreement with the numerical data is excellent, which shows that the augmented scaling theory provides a quantitative parameterization for meridional transport at both polar and mid-latitudes, where $\beta$ is significant (more extensive sweeps in parameter space are provided in Figure~\ref{fig:xi_vs_Q}). We emphasize that there are no fitting parameters left when applying the scaling theory to the inhomogeneous model: the only fitting parameter that appeared in the development of the augmented scaling theory -- the constant $2.5$ at the denominator of (\ref{fullF})\cor{, or, equivalently, the coefficient $c_\gamma$} -- was determined using the homogeneous model.


\section{Criticality of an equilibrated planetary atmosphere}

The theoretical scaling for the heat flux on a $\beta$-plane can be used to address an open question in atmospheric dynamics. Based on an analogy with convective adjustment and on observations of the thermal structure of the atmosphere, \citeA{Stone} hypothesized that planetary atmospheres equilibrate to a state of marginal stability, where the so-called criticality parameter $\xi=1/\beta_*$ is ${\cal O}(1)$. The generality of the argument  has been subsequently challenged by some numerical studies (e.g., \citeA{Panetta,Thuburn,Barry}), and it has been questioned to what extent the condition for baroclinic instability in the 2LQG model can be applied to continuously stratified fluids (e.g. \citeA{ZGL}). However the notion has regained support  through a series of numerical simulations of fully stratified atmospheres presented in \citeA{Schneider2004} and \citeA{Schneider}, which showed that the atmospheric mean state consistently equilibrates in such a way that $\xi \sim 1$ for a wide range of forcings and parameters. \citeA{Schneider2004} argued that marginal criticality in these simulations was achieved through an adjustment of the vertical stratification, a result that is precluded in QG models where the vertical stratification is imposed. Since then, some studies have reported a small but significant increase in criticality as the external forcing is increased, both in the two-layer model~\cite{ZG07} and in atmospheric models with more detailed vertical structure \cite{ZG2008,ZGV,Jansen13}. 

The present scaling theory provides a quantitative estimate of the criticality for the canonical 2LQG model. In the inhomogeneous model the criticality at $y=0$ is evaluated by considering equation (\ref{eqdytaugamma}) at $y=0$:
\begin{eqnarray}
\xi|_{y=0} = \frac{1}{c_\gamma^{\frac{1}{2+\gamma}}} \, \left(\frac{L}{\lambda} \right)^{\frac{1}{2+\gamma}} \,   \left(\frac{Q}{\beta^2 \lambda^2} \right)^{\frac{1}{2+\gamma}} \, , \label{supercritgamma}
\end{eqnarray}
valid in the (low-drag finite-$\beta$) zonostrophic regime. There is no preferred value for the criticality: it increases with the magnitude $Q$ of the imposed heat flux, albeit through a very weak power-law dependence, with an exponent $1/(2+\gamma) \simeq 0.18$. This scaling exponent is larger than the vanishing exponent associated with baroclinic adjustment, but smaller than the scaling exponent $1/4$ predicted by HL. The small value of the scaling exponent is consistent with fully stratified models which show that a rather strong increase in $Q$ induces only a modest increase in criticality. To test this prediction, we performed additional simulations of the inhomogeneous model, increasing the heat flux, for otherwise constant parameters, in a low-drag regime with $\kappa/(\beta \lambda)=0.1$ and $\lambda/L=0.02$. The emergent criticality from these numerical runs is shown in Figure~\ref{fig:xi_vs_Q}a, together with the theoretical prediction (\ref{supercritgamma}) (without fitting parameters once the value of $c_\gamma$ has been obtained from the homogeneous model). Once again, the theoretical prediction is in excellent agreement with the numerical data, both in terms of prefactor and scaling exponent. 
In Figure~\ref{fig:xi_vs_Q}b, we vary $\beta$ for otherwise constant parameters. The numerical data again fall nicely onto the full theoretical prediction associated with equation (\ref{Dsimplelin}), derived in \ref{app:Eme}. The criticality initially follows the asymptote (\ref{supercritgamma}) associated with the zonostrophic regime, before transitioning to the $\beta \to 0$ vortex-gas asymptote associated with the scaling-law (\ref{Dlinnobeta}). The $\hat{\beta}=1$ and $\hat{\beta}=3$ profiles in Figure~\ref{fig:profiles} correspond respectively to $\beta_*|_{y=0}=1/\xi|_{y=0}\simeq 0.33$ and $\beta_*|_{y=0}\simeq 0.5$. The associated values of $B$ are respectively $3.2$ and $70$, which justifies {\it a posteriori} the use of the $B\gtrsim 1$ zonostrophic scaling-law to compute the theoretical profiles displayed in Figure~\ref{fig:profiles}, for nonzero $\hat{\beta}$. The good agreement between the data points and the theoretical prediction in Figure~\ref{fig:xi_vs_Q}b illustrates the fact that the augmented scaling theory provides a convincing parameterization of heat transport in both the vortex-gas and zonostrophic regimes, but also in the crossover region between the two.

In conclusion, our results confirm that the criticality of a planetary atmosphere is weakly sensitive to changes in radiative forcing, consistent with results obtained with more complete sets of equations for fully stratified atmospheres. In particular the 2LQG model provides support for the notion that the criticality of Earth's atmosphere will remain approximately constant in response to the weak (but very important for society) changes in radiative forcing due to anthropogenic emissions of greenhouse gases.

\section{Conclusion\label{sec:Conclusion}}

\begin{figure}
\hspace{-1cm}
\includegraphics[width=15 cm]{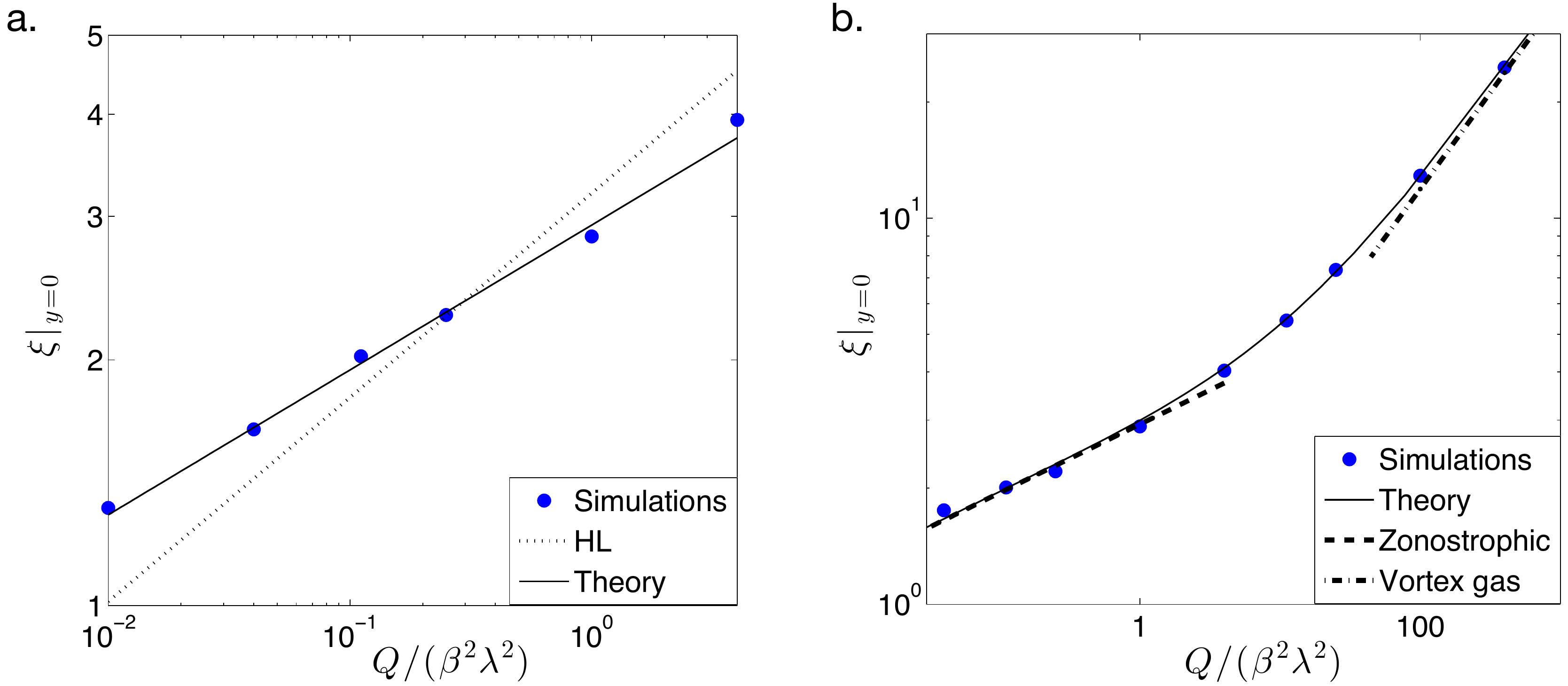} 
\caption{Emerging criticality as a function of the imposed heat flux, non-dimensionalized with $\beta^2 \lambda^2$. \textbf{a: varying $Q$ with otherwise constant parameters.} Symbols are numerical data for ${\kappa/(\beta \lambda)=0.1}$ and ${\lambda/L=0.02}$. The dotted line shows the scaling exponent $1/4$ predicted by HL, while the solid line is the theoretical prediction (\ref{supercritgamma}). \textbf{b: varying $\beta$ with otherwise constant parameters.} Symbols are DNS data for ${\hat{\kappa}=\kappa/\sqrt{Q}=0.3}$ and ${\lambda/L=0.02}$. The solid line is the theoretical prediction from the full scaling theory (derived in \ref{app:Eme}), the dashed-line at low flux is the asymptote (\ref{supercritgamma}), while the \cor{dash-dotted line} at high flux is the ${\beta=0}$ vortex-gas asymptote associated with the scaling relation (\ref{Dlinnobeta}).  \label{fig:xi_vs_Q}}
\end{figure}

Focusing on what is arguably the simplest model of baroclinic turbulence, we have derived a physically-based scaling theory for the turbulent heat transport: within this idealized framework, the scaling theory quantitatively provides the eddy diffusivity of a given patch of atmosphere or ocean as a function of the local parameters of the problem. The resulting expression of the diffusivity is given by equation (\ref{Dsimplelin}) or (\ref{Dsimplequad}), depending on whether linear or quadratic bottom drag is considered. The theory includes the crucial impact of planetary curvature within the $\beta$-plane approximation, and thus captures the mid-latitude zonostrophic regime characterized by the emergence of Rossby waves and zonal jets. \cor{The functional form of (\ref{Dsimplelin}) and (\ref{Dsimplequad}) is expected to be robust across models of baroclinic turbulence, but the fitting parameters may depend on details of the model. In other words, (\ref{Dsimplelin}) and (\ref{Dsimplequad}) can be recast into the form:}
\begin{eqnarray}
D_* & = & \frac{c_1 }{ \left[e^{-c_2/\kappa_*} +\sqrt{c_1/c_\gamma} \, \beta_*^{\gamma/2}   \right]^2} \, , \label{Dsimplelincoeffs} \\
D_* & = & \frac{c_3 }{ \left[\sqrt{\mu_*} +\sqrt{c_3/c_\gamma} \, \beta_*^{\gamma/2}   \right]^2} \, , \label{Dsimplequadcoeffs}
\end{eqnarray}
\cor{where there are three fitting parameters in the case of linear drag ($c_1$, $c_2$ and $c_\gamma$) and two in the case of quadratic drag ($c_3$ and $c_\gamma$), and $c_\gamma$ must assume the same value in both (\ref{Dsimplelincoeffs}) and (\ref{Dsimplequadcoeffs}). For $\beta_*=0$, expressions (\ref{Dsimplelincoeffs}) and (\ref{Dsimplequadcoeffs}) reduce to the predictions of the vortex gas approach in GF. The coefficients $c_1$, $c_2$ and $c_3$ were obtained in GF through simulations of the 2LQG model with $\beta=0$ and either linear or quadratic drag: expressions (\ref{Dlinnobeta}) and (\ref{Dquadnobeta}) correspond to $c_1=2.0$, $c_2=0.36$ and $c_3=2.0$. A central result of the present study is that the entire dataset for $\beta \neq 0$ and linear or quadratic drag -- i.e., both panels in Figure~\ref{fig:D_vs_beta} -- is captured by the theoretical forms (\ref{Dsimplelincoeffs}) and (\ref{Dsimplequadcoeffs}) with $c_\gamma = 0.113$. For large $\beta_*$ the diffusivity is independent of both the form and the magnitude of the drag and simply scales as $D_* \simeq c_\gamma \,   \beta_*^{-\gamma}$.} 

We then leveraged the emergent scale separation of baroclinic turbulence to turn the scaling theory into a diffusive parameterization of turbulent heat transport in an inhomogeneous system driven by a large-scale meridionally dependent external heat flux. The diffusive approach is fully justified in this context, and our predictions for the emergent meridional temperature profile are in excellent agreement with the ones obtained by numerical simulation of the turbulent flow, both in terms of shape and magnitude, see Figure~\ref{fig:profiles}.

The development of the scaling theory builds on the vortex gas theory derived in GF for $\beta=0$. The extension to nonzero $\beta$ is based on an argument put forward by HL, but adapted here to the vortex gas framework (see also \citeA{Lapeyre,CH21}). At first, as $\beta$ increases from zero, the system remains rather insensitive to $\beta$, until the Rhines wavenumber becomes comparable to the energy-containing wavenumber of the vortex gas. This is the transition point at which the system starts feeling the influence of planetary curvature: below this transition value of $\beta$, the system obeys the standard vortex-gas scaling-laws, with inverse barotropic energy transfers that saturate at a large scale determined by frictional damping. Above that transition value, the system enters a zonostrophic scaling regime. The inverse energy flux is channelled into zonal jets that do not participate in meridional transport, leading to a diffusivity that is rather insensitive to the bottom drag coefficient. This second scaling regime applies for values of $\beta$ between the transition value and the weakly nonlinear regime arising for larger $\beta$ (beyond the weakly nonlinear regime $\beta$ suppresses baroclinic instability of the undamped system, thereby suppressing any turbulence).

There have been previous studies focusing on the suppression of meridional transport by zonal jets. Some works considered broad zonal currents (aimed at modelling the full width of the Antarctic Circumpolar Current) with meridional scales much larger than the mixing-length scale~\cite{Ferrari}. The present zonostrophic regime departs from that situation, because the width of the jets is comparable to the mixing length (see also TY07). Other works assessed the importance of $\beta$ by combining Rhines' zonostrophic phenomenology with standard scaling-laws governing the diffusivity of a passive tracer in barotropic $\beta$-plane turbulence instead of starting from a vortex gas picture. \citeA{Kong} describe the barotropic turbulence with a white-noise process, an approach that has proved appropriate to describe the purely barotropic problem, but that does not seem adequate to describe the spatially and temporally coherent large-scale vortices that populate baroclinic turbulence. HL adopted a purely spectral approach, combining Rhines' arrest mechanism with the double cascade picture introduced by Salmon~\cite{Salmon80} to derive scaling predictions. That such purely spectral predictions depart from the numerical results again stems from the emergence of coherent vortices, as described in TY06. It could be that the effective kinetic energy input into the barotropic vortex gas is not localized in Fourier space, which would explain the departure between the numerical energy spectra reported by \citeA{Chang,CH21} and the spectra predicted by standard turbulent cascade arguments. Vortex dynamics leads to large-scale correlations and intense velocities near the vortex cores. At the mathematical level, this introduces a logarithmic correction in the estimation of the barotropic kinetic energy that turns out to be crucial in the $\beta=0$ scaling theory of GF. For nonzero $\beta$, the logarithmic corrections associated with the vortices keep playing a central role through logarithmic terms in the control parameter $B$. Should one ignore the logarithmic term in expression (\ref{Dvsl0}) before following the line of arguments around equation (\ref{asymptF}), it would lead to $B \sim \beta \lb0$, ${\cal F}(B)\sim B^{-2}$ and finally $D_* \sim \beta_*^{-\gamma}$ with $\gamma=2$, which is the conclusion of the purely spectral theory of HL. The discrepancy between the HL prediction $\gamma=2$ and the numerical data was pointed out early on by HL (see also Figure~\ref{fig:rescaledD}), who report a numerical exponent between $3$ and $4$, while TY07 report a value close to $4$. The present scaling approach leads to the prediction $\gamma=3.64$, providing theoretical underpinning to these observations.

This scaling exponent reflects the stiffness of the flux-gradient relation in baroclinic turbulence: the dimensional heat flux is proportional to the power $2+\gamma=5.64$ of the meridional temperature gradient. Correspondingly, the emergent gradient varies very little with the imposed flux, which is the rationale behind baroclinic adjustment. However, the present scaling theory also makes it clear that there are no preferred values of the criticality, which varies with the imposed heat flux and between various latitudes. That the criticality of zonal currents is not pinned to one is supported by the evidence that the criticality of the Antarctic Circumpolar Current in the Southern Ocean is much larger than one, even though this ocean is characterized by a reentrant baroclinically unstable current and is dynamically very similar to the midlatitude atmosphere~\cite{Tulloch}. But the changes in criticality are even more dramatic when considered on the planetary scale. Indeed, Figure~\ref{fig:xi_vs_Q}b suggests that significant variations in criticality are to be expected between different latitudes $\theta$ within the same planetary atmosphere. 
The meridional buoyancy flux $\phi_b$ of a 3D atmosphere modelled by the present two-layer setup scales as:
\begin{equation}
\phi_b = \frac{f  \la \psi_x \tau \ra}{H} \sim \frac{f \lambda^2 L Q}{H} = \frac{f \lambda^4 \beta^2 L}{H} \times \left( \frac{Q}{\beta^2 \lambda^2} \right) \, ,
\end{equation}
where $H$ denotes the layer depth, the scale $L$ of the forcing is comparable to the planetary radius, and we have used (\ref{avgeq}) to estimate the 2LQG flux $\la \psi_x \tau \ra$. Consider the situation where the same buoyancy flux $\phi_b$ is to be transferred across various latitudes $\theta$ in an atmosphere with uniform static stability and layer depths. The Rossby deformation radius $\lambda$ is then inversely proportional to $f$, and the dimensionless heating rate $Q/(\beta^2 \lambda^2)$ of the present model -- the abscissa in Figure~\ref{fig:xi_vs_Q} --  increases significantly with latitude, as $f^3/\beta^2\sim \sin^3 \theta/\cos^2 \theta$. This corresponds to an increase by a factor 16  between 30$^o$N and 60$^o$N, and a factor 160 between 20$^o$N and 70$^o$N. We thus expect a substantial increase in criticality as we move towards polar latitudes. According to Figure~\ref{fig:xi_vs_Q}b, the flow may even transition from a jet-dominated regime at low latitudes to a vortex-gas regime at higher latitudes. As a matter of fact, the image of the surface layer of Jupiter's atmosphere in Figure~\ref{fig:SO_Jupiter} points to such a transition: the banded structure extends up to latitudes of approximately 70$^{\circ}$, beyond which one observes disorganized `polar turbulence' reminiscent of the high-criticality vortex gas \cite{Porco,Sayanagi}.

Last, it should be mentioned that the scaling theory presented here ought to be further extended to make better contact with the full complexity of planetary atmospheres and oceans. A few obvious directions for future work are to consider the impact of unequal layer thicknesses, and more in general multiple layers, which has been shown to reduce the tendency toward barotropization in oceanic flows \cite{Smith2002}, the modifications introduced by large-scale topography and topographic $\beta$, with possibly an angle between the large-scale temperature gradient and the potential vorticity gradient associated with $\beta$~\cite{Arbic2004b,Arbic2004a}, the dependence on different forcing mechanisms like a surface wind stress. It would also be interesting to explore whether the quasi-geostrophic results extend to primitive-equation  solutions of fully stratified atmospheres and oceans and to test whether the extreme stiffness of the present flux-gradient relationship is maintained in such more realistic settings.

\section*{Data availability statement}
The data set is available on figshare (10.6084/m9.figshare.14546733).

\acknowledgments
\cor{We thank Niklas R\"ober for producing Figure~\ref{fig:SO_Jupiter}c and the editor Bjorn Stevens for going beyond the call of duty in helping make the paper more accessible to a wider audience and improve the quality of figures. This research is supported through the European Research Council (ERC) grant FLAVE 757239, the National Science Foundation (NSF) grant AGS-1835576 and Eric and Wendy Schmidt by recommendation of the Schmidt Futures program.}


%
%

\bibliography{beta_AGU_Advances}

%
%
%
%
%


\appendix

\section{Numerical methods\label{app:Num}}

Equations (\ref{eqq1}-\ref{eqq2}) are time-stepped inside a domain of size $[0, 2\pi L]^2$ with periodic boundary conditions, using standard pseudo-spectral methods with de-aliasing and a Runge-Kutta time-stepping scheme with adaptive time step. The linear terms are treated implicitly, except for the drag term that is treated explicitly for both linear and quadratic drag. The resolution in physical space ranges from 512$^2$ to 4096$^2$, with a domain size ranging from $2\pi L= 2\pi \times 25 \lambda$ to  $2\pi L= 2\pi \times 200 \lambda$, and hyperviscosity $\nu/(UL^7)$ typically ranging from $10^{-13}$ down to $1.2 \times 10^{-20}$.

\section{Reduction of the mixing length by $\beta$\label{app:Red}}


\begin{figure}
 \hspace{-2.3 cm}
    \includegraphics[width=18 cm]{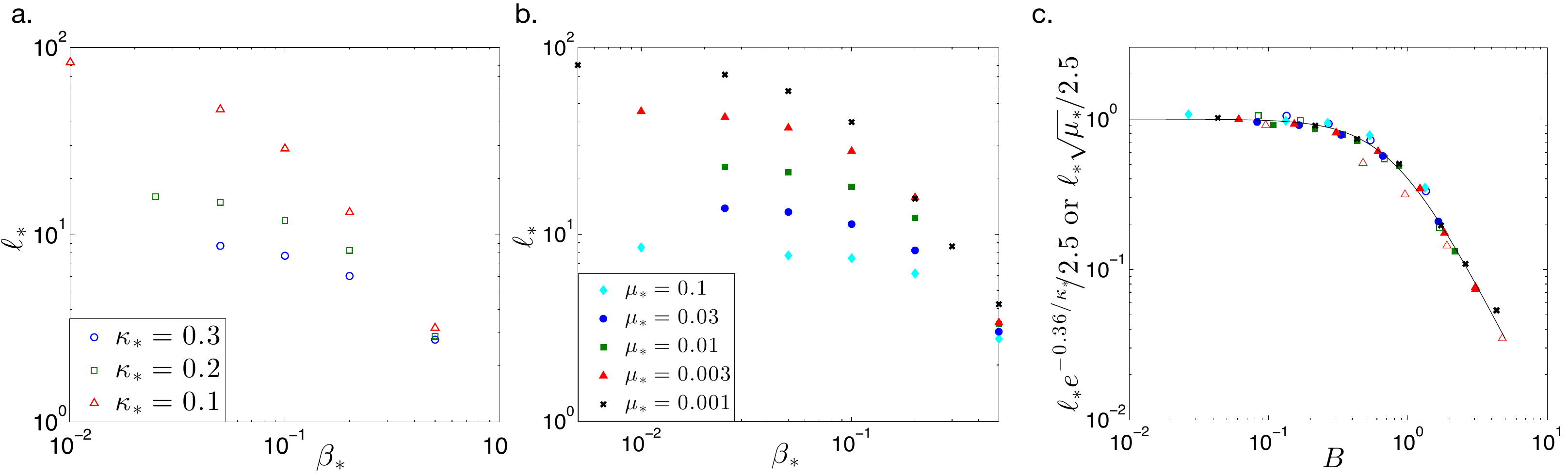} 
   \caption{\textbf{Reduction of the mixing length by $\beta$.} Diagnosed dimensionless mixing length $\ell_*$ as a function of $\beta_*$ for various values of the linear (panel a) or quadratic (panel b) drag coefficient. Panel c: the dataset collapses onto a master curve when $\ell_*/\lb0$ is plotted as a function of the control parameter $B$,  whose expression is given in equation (\ref{eq:commonB}). The theoretical expression ${{\cal G}(B)=1/(1+1.5B^{\gamma/2})}$ is shown as a solid line.
   \label{fig:ell}}
\end{figure}

The approach used to determine the scaling behavior of $D_*$ also applies to the mixing length $\ell_*$,
whose dependence on the control parameter $B$ can be written as:
\begin{eqnarray}
\ell_* = \lb0 \times {\cal G}(B) \, . \label{defG}
\end{eqnarray}
Using equation (\ref{approxlog}) to approximate the logarithmic term in $B$ and demanding that the right-hand side of (\ref{defG}) be independent of $\lb0$ for large $B$ yields the asymptotic form ${\cal G}(B) \sim B^{-\gamma/2}$, and the associated asymptotic behavior of the mixing length, $\ell_* \sim \beta_*^{-\gamma/2 \simeq -1.8}$. An alternate derivation of this scaling-law could be based on the fact that the HL scaling relation $D_* \sim \ell_*^2$ holds in both the vortex-gas and zonostrophic regimes, albeit with different prefactors, so that $D_*\sim \beta_*^{-\gamma}$ immediately yields $\ell_*\sim \beta_*^{-\gamma/2}$. A simple matching between ${\cal G}(B) \simeq 1$ for low-$B$ and ${\cal G}(B) \sim B^{-\gamma/2}$ for large $B$ is realized by the functional form:
\begin{eqnarray}
{\cal G}(B) = \frac{1}{1 + 1.5 \, B^{\gamma/2}} \, , \label{solG}
\end{eqnarray}
where the factor $1.5$ is a fitting parameter. Plots of the dimensionless mixing length are provided in Figure~\ref{fig:ell}. Once again, the control parameter $B$ leads to a very good collapse of the data onto a master curve that is well captured by expression (\ref{solG}). Inserting the functional form (\ref{solG}) into (\ref{defG}), approximating the logarithmic terms using equation (\ref{approxlog}), yields the following compact expressions for the mixing length:
\begin{eqnarray}
\ell_* & = &\frac{2.5 }{e^{-0.36/\kappa_*}+2.5\, \beta_*^{\gamma/2}} \, , \\
\ell_* & = &\frac{2.5 }{\sqrt{\mu_*}+2.5\, \beta_*^{\gamma/2}} \, , 
\end{eqnarray}
for linear and quadratic drag, respectively.


\section{Meridional temperature profile and emerging criticality \label{app:Eme}}

After integrating equation (\ref{eqdytaugamma}) over $y$, we obtain the theoretical prediction for the temperature profile in the zonostrophic regime:
\begin{eqnarray}
 \frac{\overline{\tau}(y)}{\lambda^2 \sqrt{Q}}   & = &  \hat{\beta}^{\frac{\gamma}{2+\gamma}} \left(\frac{L}{\lambda} \right)^{\frac{3+\gamma}{2+\gamma}} \frac{2+\gamma}{(3+\gamma) \, c_\gamma^{\frac{1}{2+\gamma}}}   \times \left[ F\left(\frac{1}{2}, \frac{3+\gamma}{4+2\gamma}, \frac{7+3\gamma}{4+2\gamma}, 1 \right)  \right. \label{tautruegamma}   \\
\nonumber  & & \left.  -   \cos^{\frac{3+\gamma}{2+\gamma}} \left( \frac{y}{L} \right)   \,F\left(\frac{1}{2}, \frac{3+\gamma}{4+2\gamma}, \frac{7+3\gamma}{4+2\gamma}, \cos^2 \frac{y}{L} \right)   \right] \, , 
\end{eqnarray}
where $F$ denotes the hypergeometric function. Expression (\ref{tautruegamma}) is valid for $y/L \in [-\pi/2;\pi/2]$, the temperature profile over $y/L \in [0;2\pi]$ being easily deduced from the fact that $\overline{\tau}(y)$ is symmetric to a translation by $\pi L$ accompanied by a sign change. 

One can also compute the criticality at $y=0$ as a function of the control parameters of the problem, for arbitrary $\beta$. Inserting expression (\ref{Dsimplelin}) for $D_*$ into $\lambda L Q= D_* \partial_y \overline{\tau}|_{y=0}^2$, introducing $s=  \xi \times \beta \lambda/\sqrt{Q}$, and using the approximate value $\gamma \simeq 40/11$ to alleviate notations, we obtain:

\begin{eqnarray}
\xi & = & 2.2 \left(\sqrt{\frac{2 \lambda}{L}} s - e^{-0.36 \, s/\hat{\kappa}}   \right)^{-11/20} \, , \label{xith}\\
\frac{Q}{\beta^2 \lambda^2} & = & \xi^2 / s^2 = \frac{4.85}{s^2} \left(\sqrt{\frac{2 \lambda}{L}} s - e^{-0.36 \, s/\hat{\kappa}}   \right)^{-11/10} \, . \label{Qth}
\end{eqnarray}
This is a parametric curve, the parameter being $s$, which yields the prediction for the supercriticality as a function of the control parameters. In the large-$s$ limit we recover the asymptote (\ref{supercritgamma}) in the form:
\begin{eqnarray}
\xi = 1.47 \, \left( \frac{L}{\lambda}\right)^{11/62} \left( \frac{Q}{\beta^2 \lambda^2}\right)^{11/62} \, . \label{asymptquant}
\end{eqnarray}
For large $ {Q}/({\beta^2 \lambda^2})$, we recover the $\beta$-independent vortex gas regime. With the current variables, this regime corresponds to $\xi \to \infty$, ${Q}/({\beta^2 \lambda^2}) \to \infty$, an asymptotic regime that arises when the parenthesis in (\ref{xith}) and (\ref{Qth}) vanishes. Setting this parenthesis to zero yields $s=2.78 \hat{\kappa} \, {\cal W} \left( \frac{0.255}{\hat{\kappa}}\sqrt{\frac{L}{\lambda}} \right) \simeq 1.19$ for $\hat{\kappa}=0.3$ and $\lambda/L=0.02$. The $\beta \to 0$ vortex-gas asymptote is thus $\xi \simeq 1.19 \sqrt{Q}/(\beta \lambda)$.

In Figure~\ref{fig:xi_vs_Q}b we plot the parametric curve described above, together with the asymptotes (\ref{asymptquant}) and $\xi \simeq 1.19 \sqrt{Q}/(\beta \lambda)$.

\end{document}